\documentclass[letterpaper,journal]{IEEEtran}
\usepackage{amsmath,amsfonts}
\usepackage{microtype}
\usepackage{array}
\usepackage{enumitem}
\usepackage[caption=false,font=normalsize,labelfont=sf,textfont=sf]{subfig}
\usepackage{textcomp}
\usepackage{stfloats}
\usepackage{url}
\usepackage{verbatim}
\usepackage{graphicx}
\usepackage{cite}
\usepackage{indentfirst}
\usepackage[linesnumbered,ruled,vlined]{algorithm2e}

\SetAlFnt{\footnotesize}
\SetAlCapFnt{\footnotesize}

\usepackage{algpseudocode}
\usepackage{arydshln}
\usepackage{multirow}
\usepackage{tabto}
\usepackage{url}
\usepackage{graphicx}
\usepackage{textcomp}
\usepackage{xcolor}
\usepackage{makecell}
\usepackage{tabularx}
\usepackage{pifont}
\usepackage{booktabs}
\usepackage[normalem]{ulem}
\usepackage{amssymb} % For \Delta symbol
\usepackage{titlesec}
\titleformat{\subparagraph}[runin]{\normalfont\normalsize\bfseries}{\thesubparagraph}{1em}{}{}
\titlespacing*{\subparagraph}{0pt}{3.25ex plus 1ex minus .2ex}{1em}

\titlespacing*{\section}
{0pt}{5pt}{3pt}
\titlespacing*{\subsection}
{0pt}{5pt}{3pt}
\titlespacing*{\subsubsection}
{0pt}{5pt}{3pt}
\newcommand{\ywrone}[1]{#1}

\newcolumntype{M}[1]{>{\centering\arraybackslash}m{#1}}

\newcommand{\blueyes}{\textcolor{blue}{\textbf{Yes}}}

\usepackage{array}
\newcommand{\acclist}[1]{%
    \begin{itemize}[nosep, topsep=0pt, leftmargin=*, label=\textbullet]
        \raggedright 
        #1
    \end{itemize}%
}

\newcommand{\accitem}[1]{\item #1}

\SetKwInput{Input}{Input}
\SetKwInput{Output}{Output}
\SetKw{Continue}{continue}
\SetKwComment{Comment}{$\triangleright$\ }{}

\newcounter{hypo}

\usepackage{xurl} % lets long URLs in references break at any character
\usepackage[hidelinks,breaklinks=true]{hyperref}

\begin{document}

\setlist[enumerate]{itemsep=0pt,parsep=0pt,topsep=3pt,partopsep=0pt,leftmargin=*}

% Compact float spacing; preserve original fonts and all revision colors.
\setlength{\intextsep}{3pt}
\setlength{\dbltextfloatsep}{6pt}
\setlength{\dblfloatsep}{3pt}

% \title{Gear-Ratio-Aware Standard Cell Layout Framework for 
% DTCO Exploration}

% \title{A Standard Cell Layout Framework with Enhanced Routability for DTCO Exploration}
\title{An Extended Study of Gear-Ratio-Aware Standard Cell Layout Generation for DTCO Exploration}

% \author{Chung-Kuan Cheng,\IEEEmembership{Fellow,~IEEE,}
  % <-this % stops a space
% \thanks{This paper was produced by the IEEE Publication Technology Group. They are in Piscataway, NJ.}% <-this % stops a space
% \thanks{Manuscript received April 19, 2021; revised August 16, 2021.}
% }
\author{Chung-Kuan Cheng, \IEEEmembership{Fellow,~IEEE,} 
Andrew B. Kahng, \IEEEmembership{Fellow,~IEEE,} 
Bill Lin, \IEEEmembership{Senior Member,~IEEE,} \\
Yucheng Wang, \IEEEmembership{Graduate Student Member,~IEEE,} 
and Dooseok Yoon, \IEEEmembership{Graduate Student Member,~IEEE}%
\thanks{This work was supported in part by Logic Pathfinding Lab, Samsung Semiconductor Inc., in part by NSF under Grant CCF-2110419, in part by DARPA under Grant HR0011-18-2-0032, and in part by the C-DEN Center. \textit{(Corresponding author: Yucheng Wang.)}}%
\thanks{The authors are with the University of California San Diego, La Jolla, CA 92093 USA (e-mail: yuw132@ucsd.edu).}}

\maketitle

\begin{abstract}
Advanced nodes decouple contacted poly pitch (CPP) and 
lower-metal pitch to improve routability. 
We present CPCell, an efficient standard-cell layout generation 
framework to support arbitrary gear ratio (GR) 
and offset parameters through a fine-grained 
layered grid graph and constraint-programming-based 
placement–routing co-optimization. 
Layout quality is improved via 
Middle-of-Line routing, M0 pin 
enablement, pin accessibility constraints 
and a weighted multi-objective formulation 
that jointly optimizes cell layouts. To scale to 
netlists with up to 48 transistors, we incorporate acceleration 
techniques including transistor clustering, 
identical transistor partitioning, routing 
lower bound tightening and early termination 
strategies. Comprehensive cell-level and 
block-level studies are conducted to 
evaluate GR and offset choices, quantify 
the benefits of the proposed objectives 
and assess their impact on power, performance, 
area and IR-drop outcomes.
\end{abstract}


%\vspace{-0.5cm}
\section{Introduction}
\label{sec:intro}

% While early technologies largely assumed 
% uniform scaling between contacted poly pitch (CPP) 
% and metal pitch, advanced nodes increasingly 
% decouple these dimensions to relieve routing 
% congestion on lower metal layers~\cite{regularity}. As minimum 
% metal pitch scales more aggressively than CPP~\cite{irds2022executivesummary}, 
% the relative ratio between CPP and 
% metal pitch--—commonly referred to as the 
% \emph{gear ratio} (GR)--—has emerged as an important 
% design-technology co-optimization (DTCO)~\cite{dtco} parameter 
% that directly impacts cell-level routability and 
% block-level implementation quality. 
% Reducing the M1 pitch relative to the 
% poly pitch increases the number of M1 tracks, 
% alleviating routing congestion.
Early nodes assumed uniform scaling between contacted poly pitch (CPP) and metal pitch. 
Advanced nodes decouple CPP and lower-metal pitch to 
reduce routing congestion on lower metals~\cite{regularity}. 
As metal pitch scales faster than CPP~\cite{irds2022executivesummary}, the 
CPP-to-metal-1 (M1) pitch ratio, called \emph{gear ratio} (GR), 
becomes a key design-technology co-optimization (DTCO) parameter~\cite{dtco}. 
GR affects cell-level routability and block-level implementation quality. 
Reducing the M1 pitch increases the number of M1 tracks and reduces routing congestion.
\begin{figure}[ht]
  \centering
  \includegraphics[width=0.75\columnwidth]{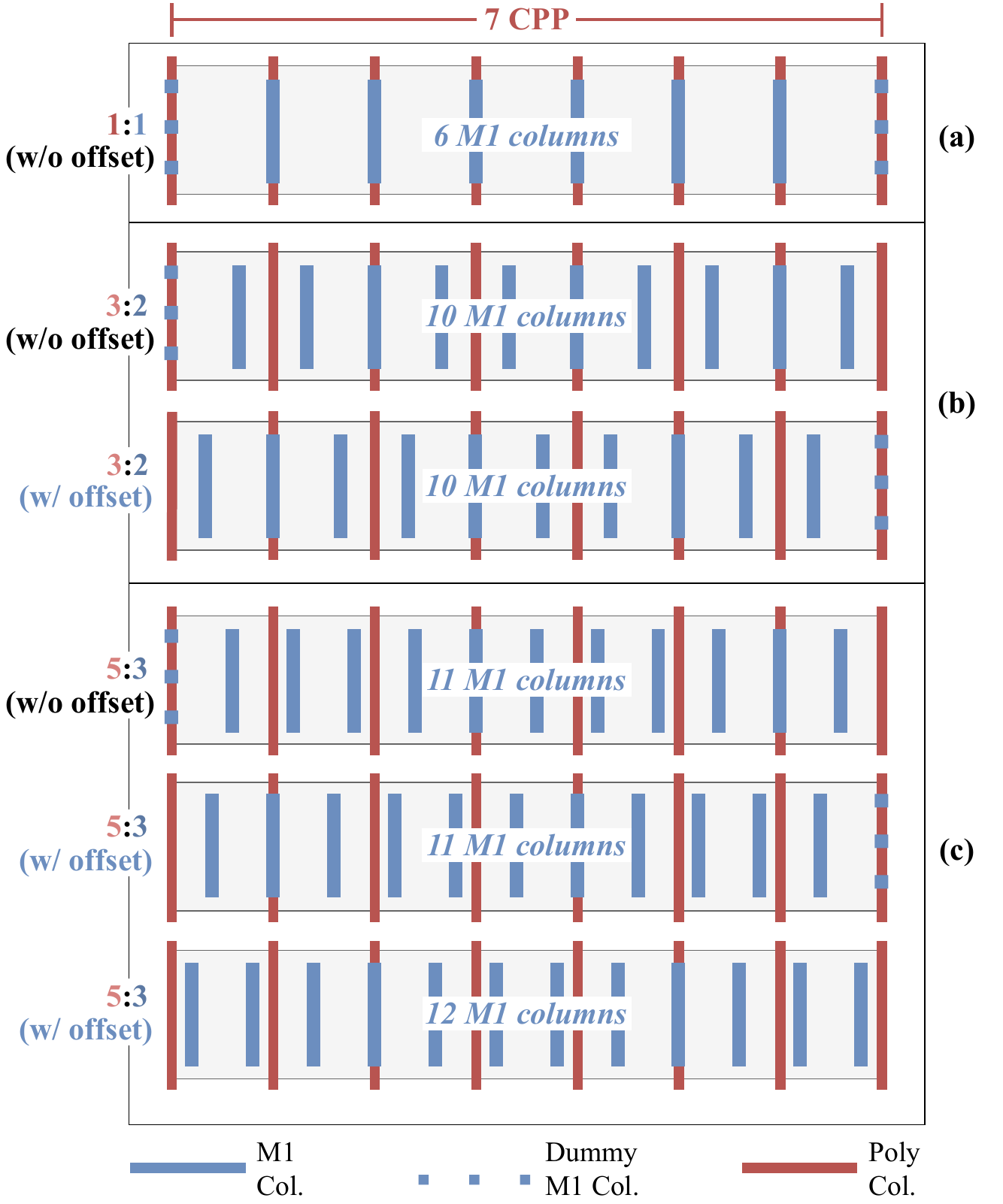}
  \caption{M1 routing resources in a 7-CPP cell for (a) 1:1 GR,
  (b) 3:2 GR and its offset variant, and (c) 5:3 GR and its two offset variants.
  Dummy columns at both cell boundaries prevent overlap with adjacent cells;
  all other columns are available for routing.}
  \label{fig:m1resource}
\end{figure}
An \emph{offset} specifies the distance between the leftmost poly and M1
columns. Shifting the M1 grid aligns local and global routing grids,
enabling placement legalization under complex GR settings. 
Figure~\ref{fig:m1resource} compares M1 routing resources within a 7-CPP cell:
1:1 GR~\cite{spr} provides six usable columns (a), 3:2 GR provides 10 (b),
and 5:3 GR provides 11 without offset or up to 12 with offset (c).
Non-1:1 GRs can misalign the poly and M1 grids~\cite{agr}; offset variants
restore alignment, preventing block-level legalization failures and enabling
accurate technology evaluations.

Prior work studies gear ratio and offset, mainly using 
post-processed layouts or backside power delivery. 
Jeong et al.~\cite{jeong2023grcells} propose a non-1:1 
gear-ratio-aware DTCO framework that improves pin accessibility 
using post-placement pin relocation and post-routing stretching. 
Their method reduces routing violations without 
increasing wirelength, timing or power. 
However, because it introduces gear ratio effects 
as post-processing, it does not model non-1:1 GR effects during cell generation. 
Cheng et al.~\cite{cheng2025invited} integrate non-1:1 gear 
ratios into SMT-based standard cell generation with 
backside power delivery, but they evaluate only area. 
Comprehensive studies in sub-7\,nm technology options~\cite{choi2023probe30} 
that jointly evaluate GR and offset choices on both 
cell-level and block-level metrics are still limited.
\ywrone{This paper extends prior work~\cite{agr}, SMTCell, with critical 
DTCO features summarized in Table~\ref{tab:dtcoitems}. CPCell inherits the
relative-layered-grid-graph placement-and-routing formulation from SMTCell~\cite{agr}\cite{smtcell}. 
We introduce CPCell with the following \ywrone{key extensions}.}
\begin{enumerate}[label=(\arabic*)]
\item \emph{CPCell Framework:} CPCell is a standard-cell 
layout generation tool and a continuation of the prior SP\&R framework~\cite{spr} 
and various other prior works~\cite{cfet}~\cite{vfet}~\cite{fastandprecise} 
with an emphasis on customization and block-level 
power, performance and area (PPA). It is compatible with CP-SAT solver from OR-Tools~\cite{ortools}, 
which employs an integer- and SAT-based encoding 
compatible with previous SMT formulations in Z3\ywrone{\cite{z3}}\cite{SMT}.
The workflow of CPCell is depicted in Figure~\ref{fig:mainflow}. 
Table~\ref{tab:dtcoitems} summarizes the design-technology co-optimization (DTCO) 
capabilities of representative standard-cell layout generation frameworks. 
Compared to prior SMT-, MILP- and heuristic-based approaches, 
CPCell is the only framework that supports
arbitrary gear ratios and offset variants under PROBE3.0
Process Design Kit (PDK)~\cite{choi2023probe30} \ywrone{when solved to exact
certification. Under the early-termination (Section~\ref{subsec:earlyterm}), 
this guarantee applies to the primary objective, cell width}.
\begin{figure}[ht]
  \centering
  \includegraphics[width=0.9\columnwidth]{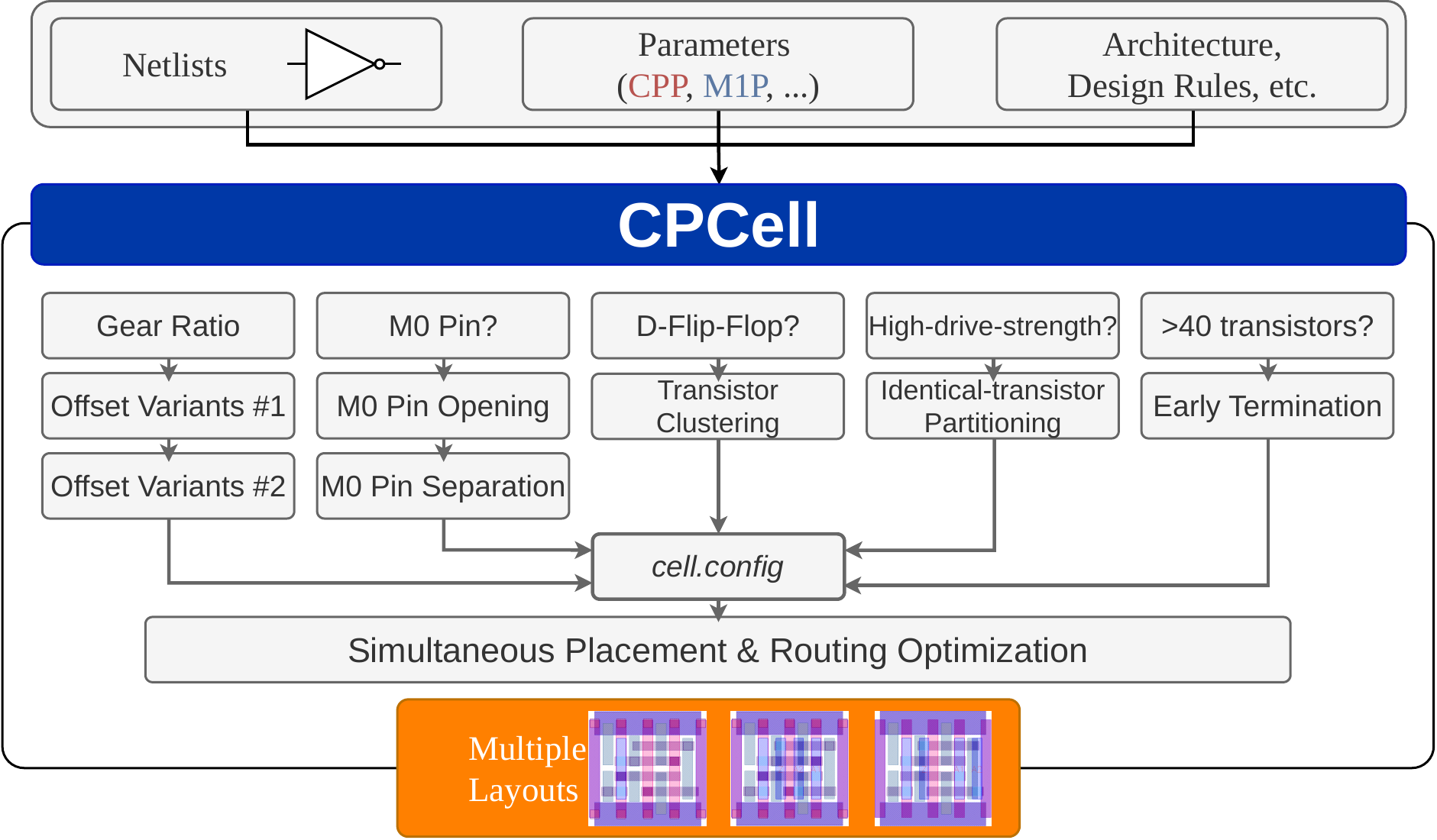}
  \caption{CPCell workflow. Inputs are the cell netlist, design rules and GR.
  Per-cell configurations specify offset variants, M0 pins and acceleration
  settings. Enumerating all offsets produces multiple layout candidates.}

  \label{fig:mainflow}
\end{figure}

\item \emph{Gear Ratio \& Offset:} Flexible metal pitch and offset
parameters modulate routing resources and cell- and block-level metrics
(Sections~\ref{sec:cell-exp} and~\ref{sec:block-exp}). Prior work~\cite{agr}
explores 3:2 and 5:3 GR without offset variants, causing legalization issues
that limit PPA. Under 3:2 GR, a 3-CPP cell aligns with the global grids
(Figure~\ref{fig:m1placement}(a)), whereas a 2-CPP cell can violate the
routing grid (b). An M1 offset restores alignment (c).
CPCell enumerates offset options for each netlist to cover all legal grid
alignments and generates the corresponding layouts.

\begin{figure*}[ht]
  \centering
  \includegraphics[width=0.88\textwidth]{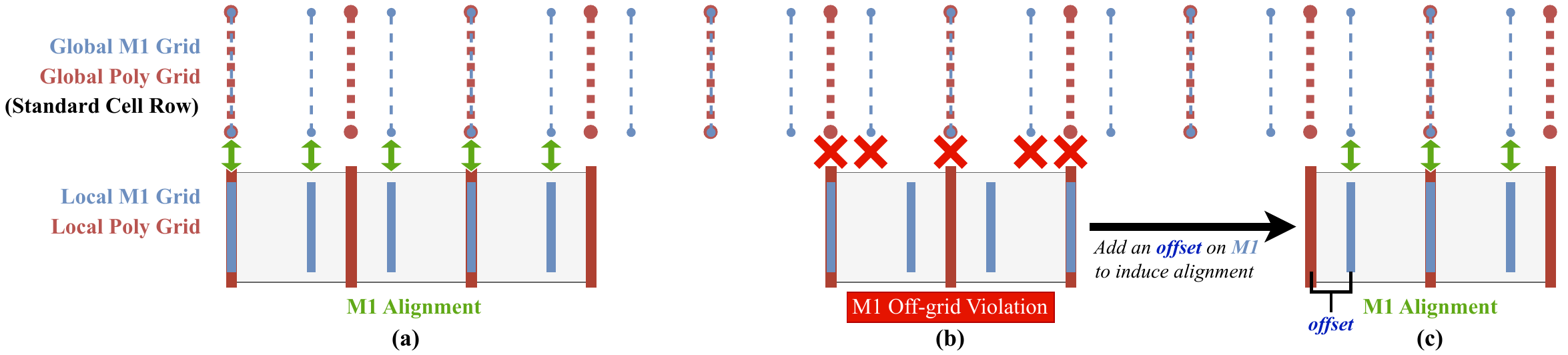}
  \caption{Placement under 3:2 GR. (a) A 3-CPP cell aligns with the global
  grids. (b) A 2-CPP cell is misaligned and requires either relocation, which burdens routing, or an M1 offset. (c) An offset aligns the 2-CPP cell at the same
  placement site as (b).}
  \label{fig:m1placement}
\end{figure*}

\item \emph{Layout enhancement:} Prior work relies 
primarily on M1 for vertical routing~\cite{agr} \cite{choi2023probe30}. 
This reliance creates routing blockages and reduces block-level routability. 
We add Middle-of-Line (MOL) routing for intra-cell 
connectivity~\cite{asap7}\cite{auMEDAL} and enable M0 pins 
to improve block-level access. 
Figure~\ref{fig:dff} compares DFFHQN\_X1 layouts under a 3:2 GR. 
CPCell reduces cell width by four CPP and enables M0 pins without adding M1 routing.
\begin{figure}[ht]
\centering
\includegraphics[width=1.0\columnwidth]{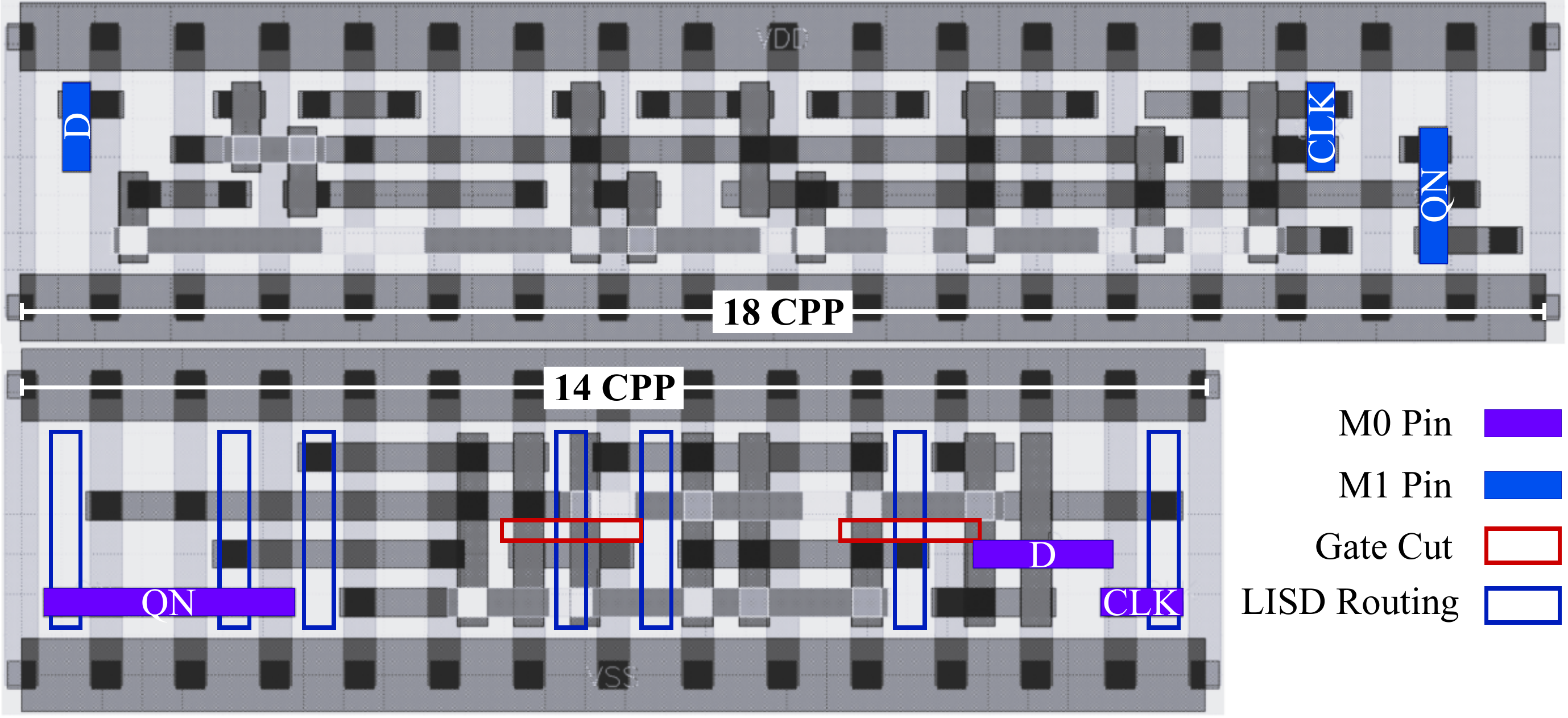}
\caption{Comparison of DFFHQN\_X1 layouts 
generated by PROBE3.0~\cite{choi2023probe30} (top) 
and CPCell (bottom) under 3:2 GR. CPCell achieves a 
4-CPP reduction with gate cuts and enables M0 
pins without M1 routing.}
\label{fig:dff}
\end{figure}

\item \emph{Acceleration Techniques:} GR and offset increase routing
complexity~\cite{agr}; for example, prior D flip-flop flows require user-given
partitioning. Our complementary techniques exploit netlist structure and
CP-SAT behavior: \emph{transistor clustering} identifies structural
regularities; \emph{identical transistor partitioning} (ITP) removes ordering
redundancy; \emph{routing lower bound tightening} (RLBT) accelerates
optimality certification; and \emph{early termination with a relative
optimality gap} trades optimality for scalability. Together, they reduce
runtime by up to 88.96\% and scale to 48-transistor netlists.

% \new{\item \emph{Equivalent Cell Group:}
% SMTCell introduces an Equivalent Cell Group (ECG) 
% flow that generates multiple layout options for 
% each logic function. Rather than producing a single 
% layout per cell, ECG systematically explores design 
% dimensions—including M1 offset variants and M0 pin 
% opening metal—to enumerate functionally 
% equivalent layouts with distinct geometric. 
% This enriched set of layout alternatives enables 
% downstream P\&R tools to select the best layout 
% instance based on the local context, improving 
% legalization robustness and block-level PPA in tight 
% gear ratio environments.}
\end{enumerate}

\begin{table*}[ht]
\centering
\caption{Summary of available DTCO features in Standard-Cell Layout Generation Tools.}
\label{tab:dtcoitems}
\renewcommand{\arraystretch}{1.15} % Adds padding to rows
\scriptsize % Reduces font size to fit page
% \begin{tabular}{|p{1.3cm}|p{1.5cm}|p{1.5cm}|p{1.4cm}|p{1.4cm}|p{1.2cm}|p{1.2cm}|p{1.4cm}|p{1.4cm}|}
\begin{tabular}{|M{1.0cm}|M{1.5cm}|M{1.5cm}|M{1.4cm}|M{1.4cm}|M{1.5cm}|M{1.1cm}|M{1.5cm}|M{1.8cm}|}
\hline
\multicolumn{1}{|l|}{\makecell{\textbf{Framework}}} 
& \textbf{SP\&R~\cite{spr}} 
& \textbf{PROBE3.0~\cite{choi2023probe30}} 
& \textbf{Csyn-fp~\cite{csynfp}} 
& \textbf{NVCell~\cite{nvcell}\cite{nvcell2}} 
& \textbf{SO3~\cite{so3cell2025}} 
& \textbf{Au-Medal~\cite{auMEDAL}} 
& \ywrone{\textbf{SMTCell-GR~\cite{agr}}} 
& \textbf{\makecell{Our Work\\(CPCell)}} \\ \hline

% \multicolumn{1}{|l|}{\textbf{Arbitrary Gear Ratio}} 
\multicolumn{1}{|l|}{\makecell{\textbf{Arbitrary}\\\textbf{Gear Ratio}}} 
& No 
& No 
& No 
& No 
& No 
& No 
& \ywrone{\blueyes} 
& \blueyes 
\\ \hline

% \multicolumn{1}{|l|}{\textbf{Arbitrary Offset}}
\multicolumn{1}{|l|}{\makecell{\textbf{Arbitrary}\\\textbf{Offset}}}
& No 
& No 
& No 
& No 
& No 
& No 
& \ywrone{No} 
& \blueyes 
\\ \hline

% \multicolumn{1}{|l|}{\textbf{Method}} 
\multicolumn{1}{|l|}{\makecell{\textbf{Method}}}
& SMT 
& SMT 
& Tree Search, DP, SMT, MILP 
& SA, RL, GA 
& MILP 
& SMT 
& \ywrone{SMT} 
& \ywrone{CP} 
\\ \hline

% \multicolumn{1}{|l|}{\textbf{Optimality Condition} } 
\multicolumn{1}{|l|}{\makecell{\textbf{Optimality}\\\textbf{Condition}}}
% & $\Delta$ (only the \newline first few \newline objectives) 
% & $\Delta$ (only the \newline first few \newline objectives)
& $\Delta$ (Prioritized) 
& $\Delta$ (Prioritized)
% & $\Delta$ (non-\newline simultaneous)
% & $\Delta$ (non-\newline simultaneous)
& $\Delta$ (Sequential)
& $\Delta$ (Sequential)
& \blueyes 
& $\Delta$ (Routing) 
& \ywrone{$\Delta$ (Prioritized)} 
& \blueyes 
\\ \hline

% \multicolumn{1}{|l|}{\textbf{MOL Routing}}
\multicolumn{1}{|l|}{\makecell{\textbf{MOL}\\\textbf{Routing}}}
& No 
& No 
& \blueyes 
& Unspecified 
& \blueyes 
& \blueyes 
& \ywrone{No} 
& \blueyes 
\\ \hline

% \multicolumn{1}{|l|}{\textbf{Parameterizable Gate Cut}} 
\multicolumn{1}{|l|}{\makecell{\textbf{Parameterizable}\\\textbf{Gate Cut}}}
& No 
& No 
& No 
& No 
& No 
& No 
& \ywrone{No} 
& \blueyes 
\\ \hline

% \multirow{18}{=}[4em]{\textbf{Acceleration \newline Technique}} 
% \multicolumn{1}{|l|}{\textbf{Acceleration Technique}}
\multicolumn{1}{|l|}{\makecell{\textbf{Acceleration}\\\textbf{Technique}}}
& \acclist{\accitem{Symmetry breaking}
          \accitem{Datapath-aware partitioning}}
& \acclist{\accitem{Symmetry breaking}
          \accitem{Datapath-aware partitioning}}
& \acclist{\accitem{Diffusion-based clustering}
          \accitem{Redundant branch pruning}
          \accitem{Fast cost estimation}}
& \acclist{\accitem{ML-based clustering}}
& \acclist{\accitem{Symmetry breaking}
          \accitem{Maximum occupying track}
          \accitem{Misaligned gate specification}
          \accitem{Structure-driven partitioning}}
& $\Delta$ (Routing)
% & \acclist{--}
& \ywrone{\acclist{\accitem{Symmetry breaking}
          \accitem{Datapath-aware partitioning}}}
& \acclist{\accitem{Symmetry breaking}
          \accitem{Transistor clustering}
          \accitem{ITP}
          \accitem{RLBT}
          \accitem{Early termination}}
\\ \hline

\multicolumn{1}{|l|}{\makecell{\textbf{Technology}}} 
& PROBE 
& PROBE 
& ASAP7~\cite{asap7}
& Unspecified 
& PROBE 
& ASAP7~\cite{asap7}
& \ywrone{PROBE} 
& PROBE \\ \hline

\end{tabular}
\begin{flushleft}
\vspace{0.1cm}
    SMT: Satisfiability modulo theories; DP: Dynamic Programming; SA: Simulated Annealing; RL: Reinforcement Learning; GA: Genetic Algorithm; MILP: Mixed Integer Linear Programming; CP: Constraint Programming; ITP: Identical Transistor Partitioning; RLBT: Routing Lower Bound Tightening. Key additions are highlighted in \textcolor{blue}{\textbf{Bold}}.
\end{flushleft} 
\vspace{-0.5cm}

\end{table*}

Section~\ref{sec:formula} introduces the proposed 
graph-based framework for constraint formulation. 
Section~\ref{sec:obj} presents the objective functions.
Section~\ref{sec:m0pin} introduces 
M0 pin accessibility constraints.
Section~\ref{sec:acc} describes the 
acceleration techniques used to reduce solver 
runtime. 
Section~\ref{sec:cell-exp} evaluates the impact 
of gear ratio and offset at cell-level using various layout 
metrics. Section~\ref{sec:block-exp} assesses 
block-level implications in terms of PPA and IR-drop.
% Our code is open-sourced on GitHub~\cite{cpcell}.
Our code, cell layout, Liberty, and evaluation scripts are 
open source and available on GitHub~\cite{cpcell}.

\section{Constraint Formulation}~\label{sec:formula}
Recall that \emph{gear ratio} (GR) is the ratio between (contacted) poly 
and M1 pitches. Common GR values include integer 
or half-integer ratios, such as 1:1 or 3:2.

Given any GR parameter, our tool dynamically 
formulates constraints based on a graph structure. 
In terms of constraint formulation, we incorporate 
largely the same set of constraints as SP\&R~\cite{spr}. 
The inherited fine-grained relative layered grid graph captures 
the gear-ratio details required for constraint formulation.
Our framework treats pitches as variables to construct 
a relative layered grid graph, representing a potential cell layout.

\begin{table}[h!]
\centering
\label{tab:symbols}
\caption{Summary of symbols and descriptions.}
% \begin{tabularx}{\columnwidth}{|c|X|} % 'X' forces the column to wrap within the remaining width
\begin{tabularx}{\columnwidth}{|p{2cm}|X|} % Changed 'l' to 'p{3cm}'
\hline
\textbf{Symbol} & \textbf{Description} \\ \hline
$G = (\mathcal{V}, \mathcal{E})$ & The relative 
layered grid graph consisting of disjoint vertex 
sets $\mathcal{V}$ and edge sets $\mathcal{E}$. \\ \hline

$v = (i; r; c)$ & A vertex triplet representing a 
specific coordinate defined by layer $i$, row $r$ 
and column $c$. \\ \hline

$L^{i}$ & The $i^{th}$ layer of the design. \\ \hline

$w_{total}, w_p, w_n,$ $c_{db}$ & The total cell width, 
PMOS transistor width, NMOS transistor width and the 
allowable diffusion break count, respectively. \\ \hline

$mp^{i}$ & The pitch value between adjacent 
tracks on layer $L^{i}$. \\ \hline

$\delta^{i}$ & The offset value from the left 
edge for layers $L^{i}$. \\ \hline
% $t$ & The number of horizontal tracks in the architecture. \\ \hline

$C^{i}$ & The column 
set for layer $L^{i}$. \\ \hline

$R^{i}$ & The row set 
for layer $L^{i}$. \\ \hline

$scx_{c}$ & The super cut node on column $c$ used 
to abstract the shape and flow of cut rows. \\ \hline

$g_{(v, \{L, R, F, B\})}$ & Geometric variables (GVs) 
assigned to vertex $v$ indicating the End-of-Line of 
a metal segment in direction left, right, front and back. \\ \hline

$P$ & The set of external pins. \\ \hline

$\mathcal{N}$ & The set of nets excluding power 
and ground nets. \\ \hline

$s_{n}$ & The source vertex of net $n$. \\ \hline

$T_{n}$ & 
The set of terminal vertices of net $n$. \\ \hline

$t_{n,k}$ & The $k^{th}$ terminal (sink) vertex of net $n$.\\ \hline

$f_{n,k}(a)$ & A binary flow variable 
indicating whether the routing path from $s_n$ 
to terminal $t_{n,k}$ uses arc $a_{(u,v)}$. \\ \hline

$\mathcal{A}$ & The set of directed arcs 
obtained by orienting each undirected edge 
in $\mathcal{E}$ in both directions. \\ \hline

$\mathcal{A}^{+}(v),\;\mathcal{A}^{-}(v)$ & The set 
of outgoing arcs from vertex $v$ and the set of 
incoming arcs to vertex $v$, respectively. \\ \hline
% (Blank line removed to avoid tabularx error)
$a_{(u,v)}$ & A directed routing arc 
from vertex $u$ to vertex $v$. \\ \hline
\end{tabularx}
\end{table}

In our graph structure, we consider a four-layer 
design\footnote{While this work utilizes a four-layer 
stack consistent with the target PDK, our proposed 
framework generalizes to arbitrary layer counts, 
subject to increased computational complexity.}: 
placement layer $L^1$ (vertical), M0 layer $L^2$ 
(horizontal), M1 layer $L^3$ (vertical), and M2 
layer $L^4$ (horizontal). Since our formulation 
relies on vertex-based constraints, accurately 
capturing the varying distances between vertices 
is crucial for satisfying complex design rules~\cite{doublepattern}.
\ywrone{Table~\ref{tab:symbols} summarizes the symbols and notation used in the formulation.}

% \vspace{-0.1cm}	
\subsection{Relative Layered Grid Graph Construction} \label{sec:graph}
Before graph construction, we determine the total cell width and layer-specific
column and row sets. The horizontal boundary $w_{\text{total}}$ depends on
the allowable diffusion break count ($c_{db}$), contacted poly pitch
($mp^1$), and accumulated PMOS ($w_p$) and NMOS ($w_n$) widths:
\begin{equation}
w_{\text{total}} = \max(w_p, w_n) + c_{db}\cdot mp^1, w_p \in \mathbb{Z^+}, w_n \in \mathbb{Z^+}.
\end{equation}

Let $i$ be the index of a layer. 
We discretize the layout area into a grid of valid 
coordinate points. These points 
correspond to the intersections of physical routing 
tracks. Let ${C}^i$ and ${R}^i$ denote 
the sets of valid x-coordinates (columns) and y-coordinates 
(rows), respectively, for each layer $L^i$. Our targeted cell layout is unidirectional, 
meaning that each layer has tracks running in a 
direction orthogonal to that of its neighbor 
layer(s), with $mp^i$ as the pitch between 
adjacent tracks. Therefore, a horizontal layer 
shares columns with its adjacent vertical 
layer(s); a vertical layer shares rows with its adjacent 
horizontal layer(s). Since gear ratio is only defined 
between vertical layers, horizontal layers must have 
the same pitch (i.e., $mp^2 = mp^4$) 
and thus all layers must share the same set of rows,

% \vspace{-0.10in}
\begin{figure}[ht]
	\centering
	\includegraphics[width=0.95\columnwidth]{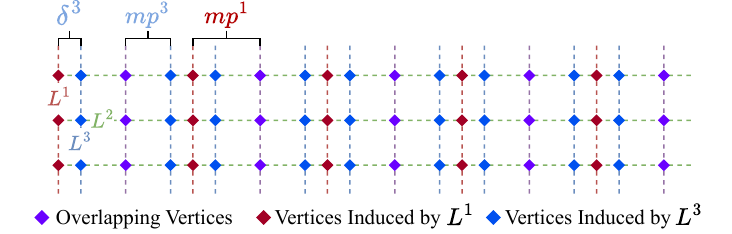}
  % \vspace{-0.4cm}
	\caption{Top-down view of the irregular pattern of columns on $L^2$
induced by different pitch values on $L^1$ and $L^3$.}
	\label{fig:2dgrid}
  % \vspace{-0.4cm}
\end{figure}

% \vspace{-0.25in}
\begin{equation}
\begin{aligned}
R^1 = R^2 = R^3 = R^4
&= \{\, h \cdot mp^i \mid h \in \mathbb{Z}^+,\, h \le t \,\}, \\
&\qquad i = 2 \text{ or } 4 .
\end{aligned}
\end{equation}
% \vspace{-0.25in}

Creating the column sets $C^i$ must allow for $mp^1 \ne mp^3$
and must consider offsets $\delta$. 
Figure~\ref{fig:2dgrid} illustrates 
how the $mp^1$ and $mp^3$ 
values can result 
in an irregular pattern of columns on $L^2$. 
The column set $C^2$ must 
contain all columns from both $C^1$ and $C^3$ 
to represent the densely 
placed routing tracks, with overlapped columns 
being merged to avoid redundancy. 
We achieve this using 
Algorithm~\ref{alg:c_iter}, which iteratively 
creates columns for vertical layers first (Lines 1--7) 
and then merges them on 
horizontal layers without repetition (Lines 8--10), ensuring 
uniqueness for each column in a horizontal layer.

\setlength{\intextsep}{0pt} 
\begin{algorithm}[tb]
\caption{Column Set Creation}
\label{alg:c_iter}
\SetKwProg{proc}{Procedure}{}{end}
\SetKwData{}{left}\SetKwData{This}{this}\SetKwData{Up}{Up}
% // iterate through vertical layers\\
\For{$i \in \{1, 3\}$}{
    $c \leftarrow \delta^i$\\
    $C^i \leftarrow \emptyset$\\
    \While{$c \le w_{\text{total}}$} {
        $C^i \leftarrow C^i \cup \{c\}$\\
        $c \leftarrow c + mp^i$
    }
    $\mathcal{C} \leftarrow C \cup C^i$
}
% // iterate through horizontal layers\\
\For{$i \in \{2, 4\}$}{
    $C^i \leftarrow C^{i-1} \cup C^{i+1}$\quad\quad// Implicitly, $C^5 = \emptyset$\\
    $\mathcal{C} \leftarrow \mathcal{C} \cup C^i$
}
\Return $\mathcal{C}$ 
\end{algorithm}

After creating $\mathcal{C}$ and $\mathcal{R}$, 
we construct a \textit{relative layered grid graph} 
$G=(\mathcal{V}, \mathcal{E})$, 
where $\mathcal{V}$ and $\mathcal{E}$ are a set of 
disjoint vertex sets and a set of 
disjoint edge sets, respectively:
\begin{align}
  \mathcal{V} &= \{V^i|i\in\{1, ..., 4\}, \\
  &\quad V^i = \{ v| v = (i; r; c), r \in R^i, c \in C^i\}\}\mathrm\, \displaybreak[1] \\
  \mathcal{E} &= \{E^i|i\in\{1, ..., 4\}, \\
  &\quad E^i = \{ e| e = (v_1, v_2),\; v_1 \in V^i, v_2 \in V^j, j\in\{i-1, i\}\}\}. 
\end{align}
% All vertices in $\mathcal{V}$ are created by the crossing points 
% between the columns and rows on each layer and are 
% indexed by layer $i\in\{1, ..., 4\}$, $row\in R^i$, and $col \in C^i$.
Each vertex $v$ is a triplet $(i;r;c)$ that contains layer $i$, 
row $r$ and column $c$. 
Each edge $e$ is a pair of vertices 
whose locations are adjacent (i.e., nearest 
neighbors) created along the direction of this layer. 
% Figure~\ref{fig:Routing} shows such an example: each colored square 
% is a vertex and each colored arrows for horizontal layers, 
To generate $\mathcal{V}$ and $\mathcal{E}$, we 
use Algorithm~\ref{alg:graph_iter}: 
for horizontal layers, edges are created along 
each row (Lines 8--10); for vertical layers, edges are 
created along each column (Lines 11--13). 
These edges facilitate intra-layer routing. Edges between layers, 
called vias, are created if and only if the 
two vertices are on adjacent layers and have 
the same $row$ and $column$ (Lines 14--17).

\setlength{\intextsep}{0pt}
\begin{algorithm}[tb]
\caption{Relative Layered Grid Graph Construction}
\label{alg:graph_iter}
\SetKwProg{proc}{Procedure}{}{end}
\SetKwData{}{left}\SetKwData{This}{this}\SetKwData{Up}{Up}
\For{$i \in \{1 ... 4\}$}{
  $V^i \leftarrow \emptyset$, $E^i \leftarrow \emptyset$\\
  // In the order of increasing rows and columns
  
  \For{$r \in R^i$}{
  \For{$c \in C^i$} {
  $v \leftarrow (i; r; c)$\\
  $V^i \leftarrow V^i \cup \{v\}$\\
  \If{layer $i$ is horizontal} {
  Get the nearest left vertex $v_{a}$ in $V^i$\\
  $E^i \leftarrow E^i \cup \{\{v, v_{a}\}\}$
  }
  \If{layer $i$ is vertical} {
  Get the nearest lower vertex $v_{b}$ in $V^i$\\
  $E^i \leftarrow E^i \cup \{\{v, v_{b}\}\}$\\
  }
  // if via can be constructed below\\
  $v_{c} \leftarrow (i - 1; r; c)$\\
  \If{$v_{c} \in V^{i-1}$} {
  $E^i \leftarrow E^i \cup \{\{v, v_{c}\}\}$
  }
  }
  }
  \ywrone{$\mathcal{V} \gets \mathcal{V} \cup V^i$, $\mathcal{E} \gets \mathcal{E} \cup E^i$}
}
\Return $\mathcal{V}$, $\mathcal{E}$
\end{algorithm}
% \vspace{-0.4cm}

\subsection{Routing}\label{sec:mcf}
\begin{figure}[ht]
  \centering
  \includegraphics[width=1.0\columnwidth]{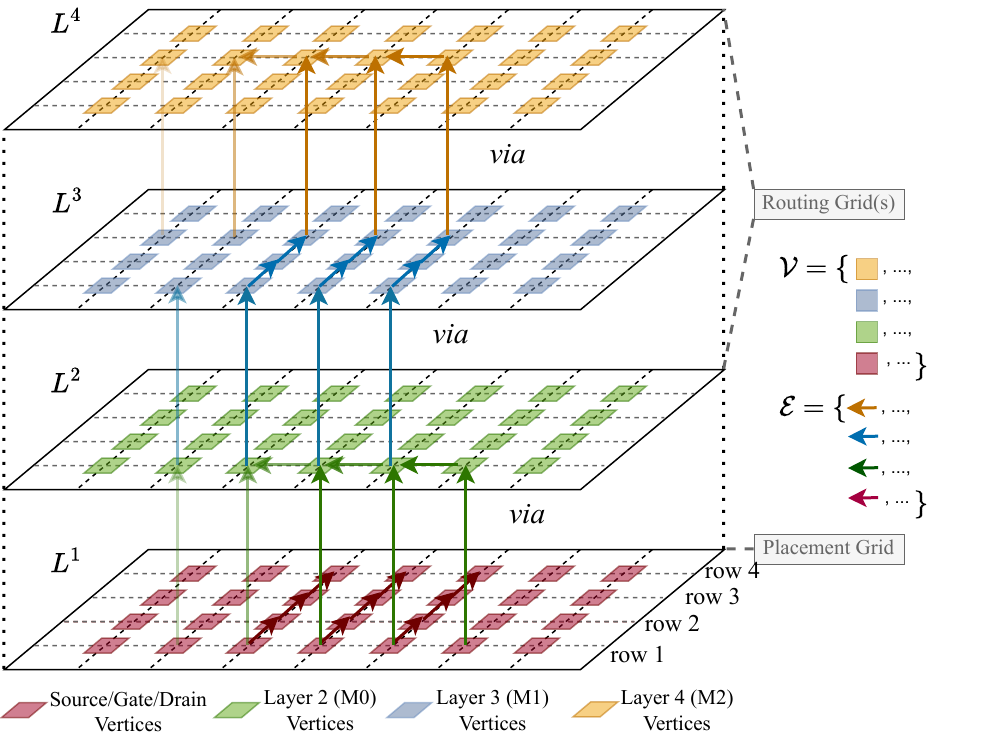}
  % \vspace{-0.4cm}
  \caption{Routing source/drain and gate terminals to frontside IO
  through $L^2$--$L^4$. Super cut nodes $scx$ control middle-row flow on each
  placement-grid column. Intra-layer connections follow layer orientation;
  vias connect layers at access points (colored squares).}
  \label{fig:mcmf}
\end{figure}

We model routing as a \emph{multi-commodity flow}
problem~\cite{HanKL2015}~\cite{Hu1963} on the relative layered grid graph
(Figure~\ref{fig:mcmf}). CPCell's directed encoding empirically scales
better than prior undirected models \cite{agr}\cite{fastandprecise}.
Each edge $\{u,v\} \in E$ becomes two directed arcs
$a_{(u,v)}, a_{(v,u)} \in \mathcal{A}$; flow is defined exclusively on
these arcs to control direction and capacity.

\textbf{Middle-of-Line (MOL) Routing.}
MOL routing~\cite{asap7} uses the Local Interconnect Source-Drain (LISD) 
and gate layers to connect adjacent terminals without 
consuming (global) metal routing resources. When terminals align in 
a column, they can be ``merged'' through the shared 
active region. This allows routing to treat the column 
as a single sink, rather than requiring distinct flows for every terminal.

Let $\mathcal{N}$ denote the set of signal nets. For each net 
$n \in \mathcal{N}$, let $s_n$ be the source and 
$T_n = \{t_{n,0}, \ldots, t_{n,K_n-1}\}$ be the set of sink terminals. 
We model global routing using binary flow variables:
\begin{equation}
f_{n,k}(a) \in \{0,1\}, \quad \forall a \in \mathcal{A},
\end{equation}
indicating whether the path from $s_n$ to terminal $t_{n,k}$ traverses 
arc $a$. To capture the interaction between global routing flow and local MOL 
connectivity, we introduce \emph{super cut node}, 
$\mathit{scx}_c \in \{0,1\}$, which determines the physical continuity 
of the placement grid ($L^1$) at column $c$. $\mathit{scx}_c = 1$ 
represents a physical break---either as a gate cut at even columns 
or a LISD cut at odd columns---which enforces electrical isolation and 
requires terminals to be routed individually by metal layers. Conversely, 
$\mathit{scx}_c = 0$ implies a continuous active region where terminals 
are electrically shorted. The local connectivity allows the router 
to bypass flow requirements for individual terminals, provided 
the column itself is connected. 
$\mathit{scx}_c$ is defined as an abstract 
cut value associated with column $c$.

\begin{figure}[ht]
	\centering
	\includegraphics[width=0.95\columnwidth]{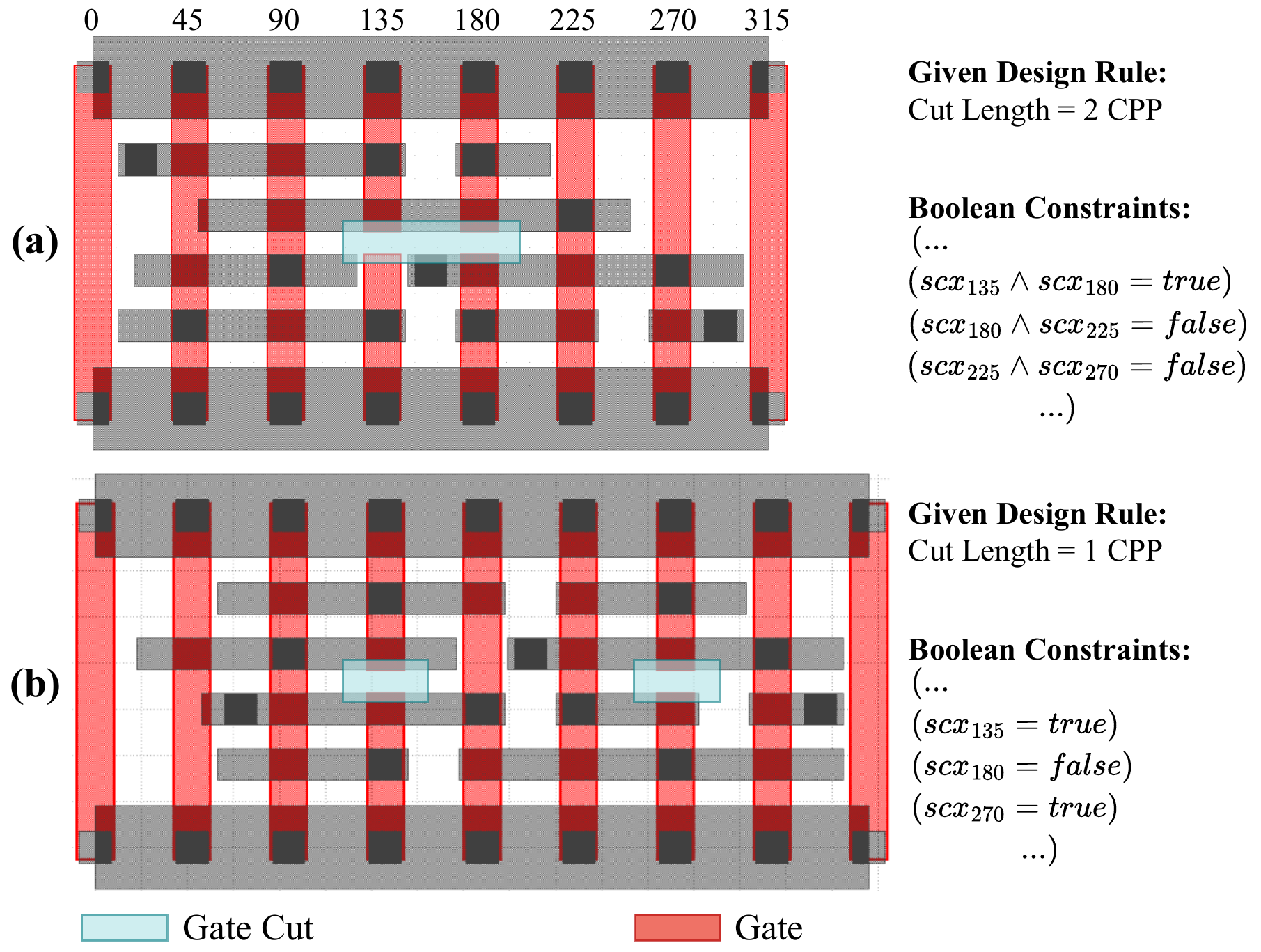}
  % \vspace{-0.4cm}
	\caption{MUX2\_X1 with (a) 2-CPP and (b) 1-CPP cut widths.
    Super cut nodes $scx$ abstract cut shapes: adjacent columns
    activate together in (a), while nodes activate individually in (b).}
	\label{fig:gc}
  % \vspace{-0.1cm}
\end{figure}

For gate cuts, the \emph{cut width} is specified in contacted poly pitch
(CPP) units~\cite{gatecut}. We enforce it by grouping super cut nodes
across adjacent even columns (Figure~\ref{fig:gc}); a two-CPP cut requires
paired activation of $scx_c$ on neighboring even columns. Cut width affects
cell width: Figure~\ref{fig:gc}(a), with a 2-CPP cut, uses one fewer CPP
than Figure~\ref{fig:gc}(b), with a 1-CPP cut. Unless specified otherwise,
we use 2-CPP cuts to reduce the area of cross-coupled structures in
D flip-flops, latches and multiplexers.

\textbf{Flow Conservation Constraints.}
For each terminal $t_{n,k}$ of net $n$ that is 
not merged by MOL routing, the flow must satisfy the following conservation laws:
\begin{equation}
\sum_{a \in \mathcal{A}^{+}(v)} f_{n,k}(a) 
- \sum_{a \in \mathcal{A}^{-}(v)} f_{n,k}(a) = 
\begin{cases} 
1 & \text{if } v = s_n \\
-1 & \text{if } v = t_{n,k} \\
0 & \text{otherwise}
\end{cases}.
\end{equation}

\subsection{Layer-based Fine-grained Design Rule Checking} \label{sec:dr}
CPCell enforces the complete set of 
conditional design rules
specified by the target PDK~\cite{choi2023probe30}, 
including Minimum Area Rule (MAR), Via Separation, End-Of-Line spacing (EOL), 
Step Height Rule (SHR) and Parallel Run Length (PRL). We parameterize 
these rules per layer and encode them using geometric 
variables to ensure full compliance.

Geometric variables are introduced in \cite{spr} to indicate 
the End-Of-Line of a metal segment. 
\begin{figure}[ht]
	\centering
	\includegraphics[width=1.0\columnwidth]{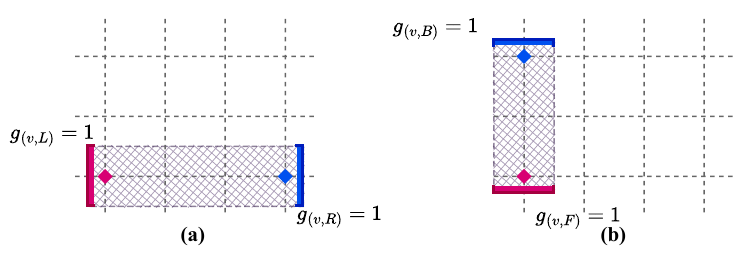}
  % \vspace{-0.4cm}
	\caption{Vertex geometric variables mark metal endpoints in four
 directions: (a) left/right and (b) front/back.}
	\label{fig:gv}
  % \vspace{-0.1cm}
\end{figure}
We define geometric variables (GVs) in our relative layered grid graph 
by assigning a set of Boolean variables to each vertex 
$v$: ${g_{(v, L)}, g_{(v, R)}, g_{(v, F)}, g_{(v, B)}}$, representing 
the four directions—left, right, front and back.\footnote{Front 
corresponds to the bottom endpoint (lower row index) and 
Back corresponds to the top endpoint (higher row index).} 
On horizontal 
layers, a left-end GV $g_{(v, L)}$ must be paired 
with exactly one right-end GV $g_{(u, R)}$ to create a 
horizontal metal segment (see Figure~\ref{fig:gv}(a)).
Similarly, on vertical layers, a front-end GV $g_{(v, F)}$ must 
be paired with one back-end GV $g_{(u, B)}$ to create 
a vertical metal segment (see Figure~\ref{fig:gv}(b)).

The non-uniformity in a relative layered grid 
graph requires the design rule parameters to be defined 
in terms of Euclidean distance. 
Prior methods~\cite{spr}~\cite{fastandprecise}~\cite{road} 
define these design rule 
parameters in terms of ``number of vertices'', which is 
valid only in a uniform grid where all edge lengths 
are equal. 
% \vspace{0.1cm}
\begin{figure}[ht]
	\centering
	\includegraphics[width=1.0\columnwidth]{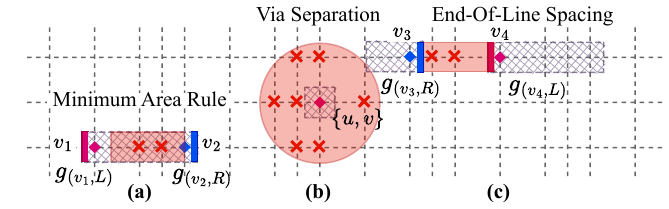}
 	\caption{Design rule checking on $L^2$ with 
    distance input. The red-shaded region represents 
    the prohibited area. (a) Minimum Area Rule (MAR) 
    uses geometric variables and prohibits the current 
    metal segments from ending too closely together. 
    (b) Via Separation uses edge variables and prohibits 
    any adjacent via within the Euclidean distance. (c) End-Of-Line 
    spacing (EOL) uses geometric variables and prohibits any 
    adjacent metal from ending too close to the current segment. 
 }
	\label{fig:eol}
  % \vspace{-0.1cm}
\end{figure}

We input design rule parameters based on distance and track 
the total distance traveled along each path until the desired 
distance is attained. Design rule parameters can also be defined 
for each layer, as different layers may have varying 
metal pitches.
Figure~\ref{fig:eol} demonstrates three of the design 
rules: MAR, Via Separation and EOL.
Figure~\ref{fig:eol}(a) shows the usage of geometric variables to 
encode the minimum area of a metal segment. 
Once $g_{(v_2, R)}$ is instantiated, the leftward 
$g_{(v_1, L)}$ within the user-defined length is 
prohibited (indicated by the red cross sign). 
Similarly, Figure~\ref{fig:eol}(c) shows the usage 
of geometric variables 
to encode the end-of-line spacing of a metal segment. 
Once $g_{(v_4, L)}$ is instantiated, the leftward 
$g_{(v_3, R)}$ within the user-defined length is 
prohibited (indicated by the red cross sign). 
Via Separation, on the other hand, does not use any 
geometric variables as illustrated in Figure~\ref{fig:eol}(b). 
Each via is represented by an undirected edge $\{u, v\}$, 
where $u=(i-1;r;c)$ and $v=(i;r;c)$. Once a via edge $\{u, v\}$ 
is instantiated, all vias within the user-defined radius 
are prohibited (indicated by the red cross sign). We encode these three rules using the Boolean 
cardinality constraint at-most-1 from CP-SAT. 
For other design rules, we examine all routing edges and their neighboring
vertices to prohibit access within the specified distance.

\section{Multi-Objective Optimization}~\label{sec:obj}
We formulate cell layout generation as 
a weighted multi-objective optimization problem 
solved using CP-SAT~\cite{ortools}. Prior 
methods~\cite{spr}\cite{agr} apply 
lexicographic optimization, which minimizes 
objectives sequentially and may sacrifice 
optimality beyond the first few objectives. 
In contrast, we guarantee 
global optimality for the weighted-sum objective 
\ywrone{when solved to exact certification}. 
While cell width and wirelength are sufficient 
to optimize primary cell-level metrics, 
\emph{auxiliary objectives} play a critical role 
in steering the solver toward the optimal 
solution more efficiently. 
The complete set of objectives is defined as follows.

\begin{enumerate}[label=(\arabic*)]
    \item $CW$: Cell width measured 
    in the number of contacted poly pitch (\#CPP).
    \item $SGD$: Total \#source/gate/drain shared.
    \item $WL$: Weighted wirelength.
    \item $DBX$: Diffusion break placement, measured by 
    the sum of placement column indices of all diffusion breaks. \label{obj:dbx}
    \item $M2$: Top metal (M2) utilization, defined as the count of M2 tracks containing at least one active routing segment. 
\end{enumerate}

\begin{figure}[ht]
	\centering
	\includegraphics[width=1.0\columnwidth]{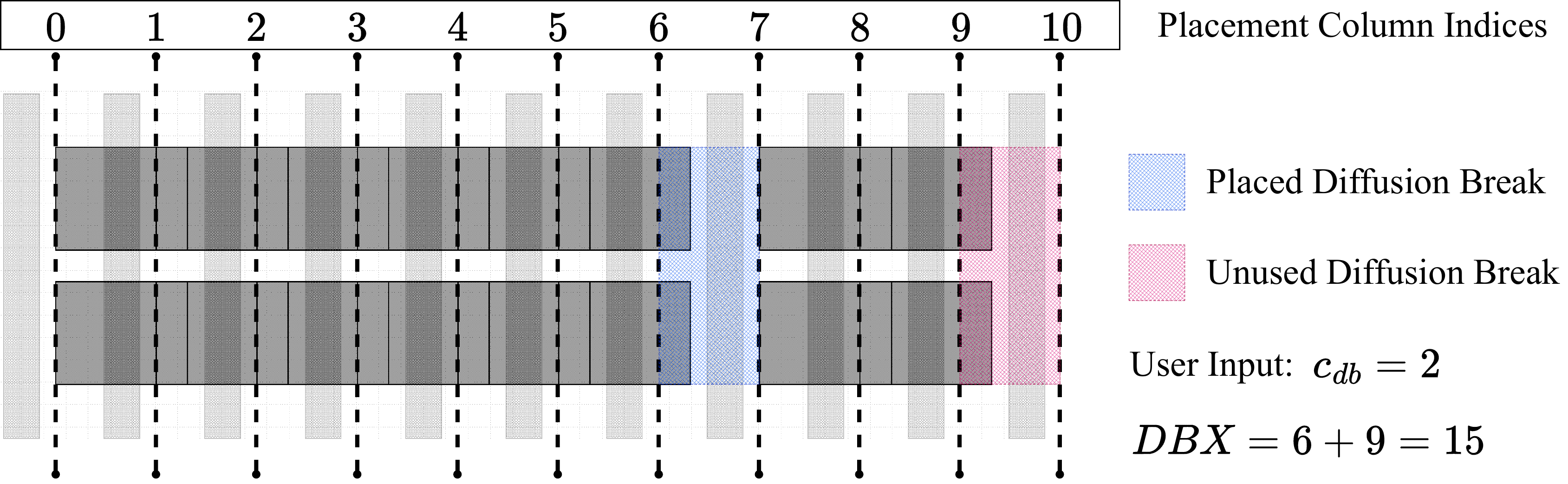}
	\caption{Illustration of diffusion break placement 
    and the $DBX$ objective. When the maximum allowable 
    diffusion breaks ($c_{db}$) exceed the layout 
    requirements, the objective forces surplus 
    breaks (pink) to the far right edge, minimizing 
    the active cell width.}
	\label{fig:dbx}
\end{figure}

Together, the final objective function is: 
\begin{equation}
\label{eq:multiobj_unnorm}
\begin{aligned}
\min
\quad
& \lambda_0 \cdot CW
+ \lambda_1 \cdot WL
- \lambda_2 \cdot SGD \\
& - \lambda_3 \cdot DBX
+ \lambda_4 \cdot M2.
\end{aligned}
\end{equation}
Unless stated otherwise, we use fixed integer 
weights in all experiments: $\lambda_0 = 1000$ and 
$\lambda_1 = \lambda_2 = \lambda_3 = \lambda_4 = 1$. We choose 
$\lambda_0$ to make $CW$ the primary optimization target.
$SGD$, $DBX$ and $M2$ are auxiliary objectives that bias 
the search towards favorable structural patterns—such as 
increased MOL routing, rightward diffusion break placement
and reduced M2 usage—thereby accelerating 
convergence without altering the optimal $CW$ and $WL$.
The runtime impact of each auxiliary objective is 
detailed in Section~\ref{subsec:ablation} via an ablation study. 

While the roles of $SGD$ and $M2$ are 
relatively intuitive, the purpose of $DBX$ is less 
obvious. Figure~\ref{fig:dbx} shows an example of 
diffusion break placement and the $DBX$ objective value.
The $DBX$ objective encourages diffusion 
breaks to be placed toward the right boundary of 
the layout canvas. Importantly, the user-defined 
parameter $c_{db}$ serves only as an \textit{upper bound} 
on the available diffusion breaks; CPCell 
automatically determines the actual number required 
for a layout. To ensure satisfiability, $c_{db}$ 
is often set higher than the minimum necessary count. 
In these cases, the $DBX$ objective naturally pushes 
any surplus (unused) diffusion breaks to the far 
right edge, preventing them from fragmenting the active 
region or increasing the effective cell width. Consequently, 
maximizing the rightward placement of diffusion breaks implicitly 
minimizes the occupied layout area, reinforcing the 
primary cell width objective.

\section{M0 Pin Enhancement}\label{sec:m0pin}
\begin{figure}[ht]
	\centering
	\includegraphics[width=0.88\columnwidth]{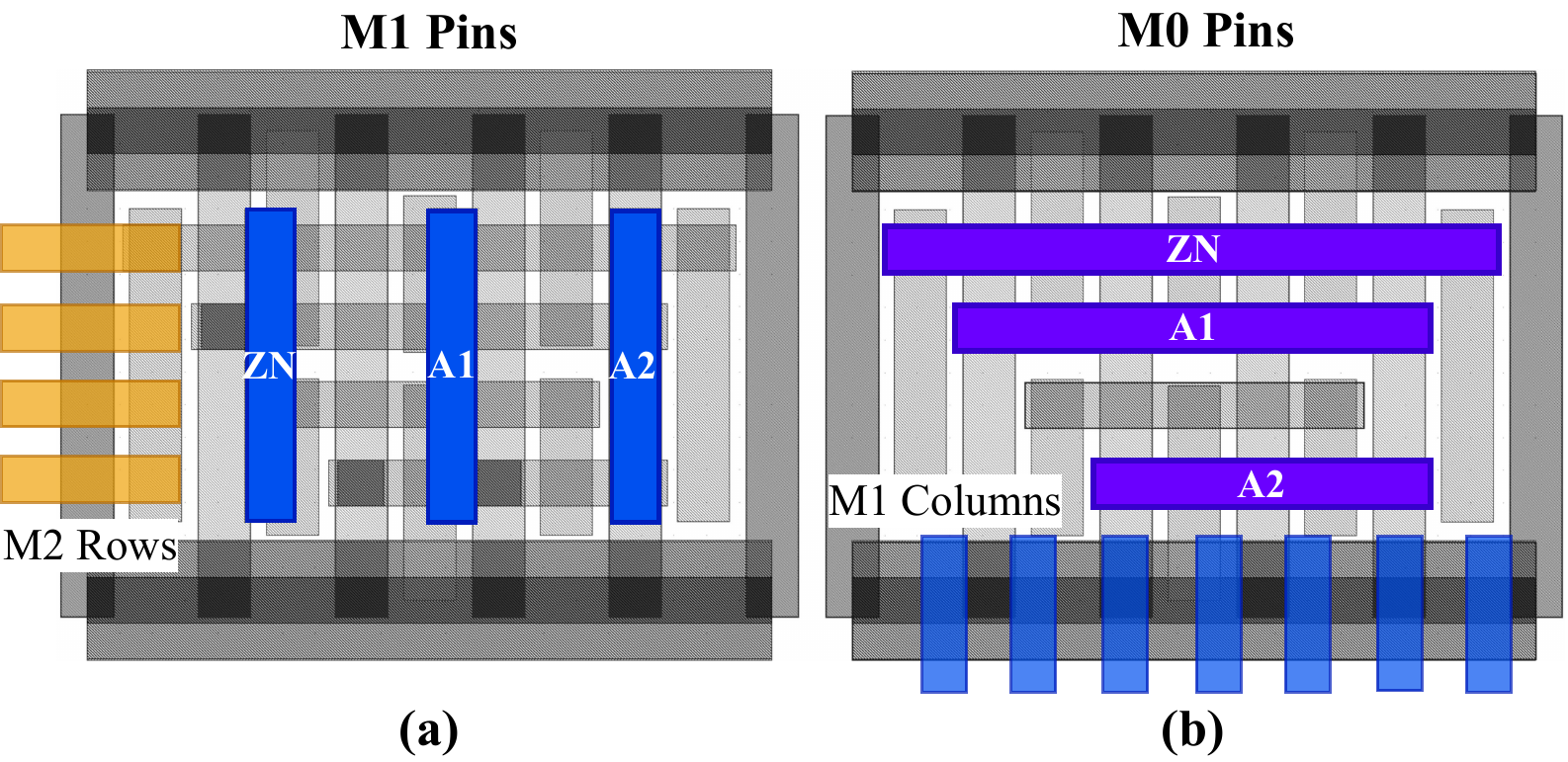}
  % \vspace{-0.4cm}
	\caption{NAND2\_X2 pin access under 3:2 GR: (a) M1 pins accessed by
M2 rows (golden yellow); (b) M0 pins accessed by M1 tracks (dark blue).}
	\label{fig:m0pin}
  % \vspace{-0.1cm}
\end{figure}

With enabled MOL routing, 
many nets no longer require M1 for 
internal routing, allowing M1 pins to be 
replaced by M0 pins and thereby increasing the 
number of access points available to the block-level 
router (see Figure~\ref{fig:m0pin}). Within the given cell width,
an M1 pin is accessible through four M2 routing tracks, an 
M0 pin can overlap up to seven M1 routing tracks, 
and it further benefits from additional routing  
resources enabled by non-1:1 gear ratios at the 
M1 layer. \ywrone{We follow 
the same set of constraints as SP\&R~\cite{spr} to guarantee
pin accessibility at M1.}
However, M0 pin accessibility is 
limited by routing blockages at both the M0 
and M1 layers: pin-to-pin and pin-to-metal spacing 
rules limit extensibility at M0, while internal 
M1 metals may fully block routing tracks above. 
In CPCell, these effects are captured through
two hard constraints--pin separation
(PS) and minimum pin opening (MPO)--which together
guarantee robust pin accessibility.

\subsection{Pin Separation (PS)}
\label{sec:pin-separation}
Let $\mathcal{P}$ denote the set of external pins. 
With M0 pin enablement, each pin $p \in \mathcal{P}$ 
is assigned at least one access vertex on the M0 layer.
We denote the representative access vertex of pin $p$ by
\begin{equation}
v_p = (2; r_p; c_p) \in V^2.
\end{equation}
If the number of available M0 rows is at least 
$|\mathcal{P}|$, CPCell enforces PS as a 
hard constraint by requiring each pin to occupy a distinct M0 row:
\begin{equation}
r_{p_i} \neq r_{p_j}, \quad \forall\, p_i \neq p_j,\; p_i, p_j \in \mathcal{P}.
\end{equation}
PS isolates pin access points across rows, complementing minimum pin
opening (MPO), which expands access regions along M0.

\subsection{Minimum Pin Opening (MPO)}
\label{sec:minimum-pin-opening}
While PS enforces spatial isolation 
among pins, each external pin must
also expose a sufficient number of access points 
to adjacent M1 routing tracks 
to ensure block-level routability. Minimum Pin Opening 
(MPO) enforces a
hard lower bound on the horizontal extent of each M0 pin, 
measured by the
number of accessible M1 routing tracks it overlaps.

For pin $p \in \mathcal{P}$, let $\mathcal{C}_p \subseteq C^2$ be the
unoccupied M1 tracks overlapping the pin. CPCell requires the hard constraint:
\begin{equation}
|\mathcal{C}_p| \ge \Theta, \quad \forall\, p \in \mathcal{P},
\end{equation}
where $\Theta$ is the user-specified minimum pin opening.
Each external pin thus exposes multiple access points despite internal
M1 routing blockages. Together, PS and MPO provide spatial isolation and
sufficient pin exposure for downstream routing under congestion.

\section{Acceleration Techniques}~\label{sec:acc}
Gear ratios and offsets increase routing resources and search
complexity~\cite{agr}. SP\&R partitioning requires expert knowledge of
optimal datapath structures~\cite{dthesis}; SMTCell-GR pre-partitioning
reuses solutions across related netlists~\cite{agr}; and
SO3~\cite{so3cell2025} uses structural partitioning, misaligned gate
specification and maximum occupying track. Their reliance on manual
intervention or prior design knowledge limits application to unseen
netlists. We introduce four complementary techniques: \textit{transistor
clustering} groups strongly coupled devices to reduce placement
granularity; \textit{identical transistor partitioning} (ITP) removes
equivalent permutations through canonical ordering; \textit{routing
lower bound tightening} (RLBT) accelerates optimality certification; and
\textit{early termination} uses a relative gap to trade optimality for
runtime on large netlists.

\subsection{Transistor Clustering} \label{subsec:cluster}

\begin{figure}[ht]
  \centering
  \includegraphics[width=0.95\columnwidth]{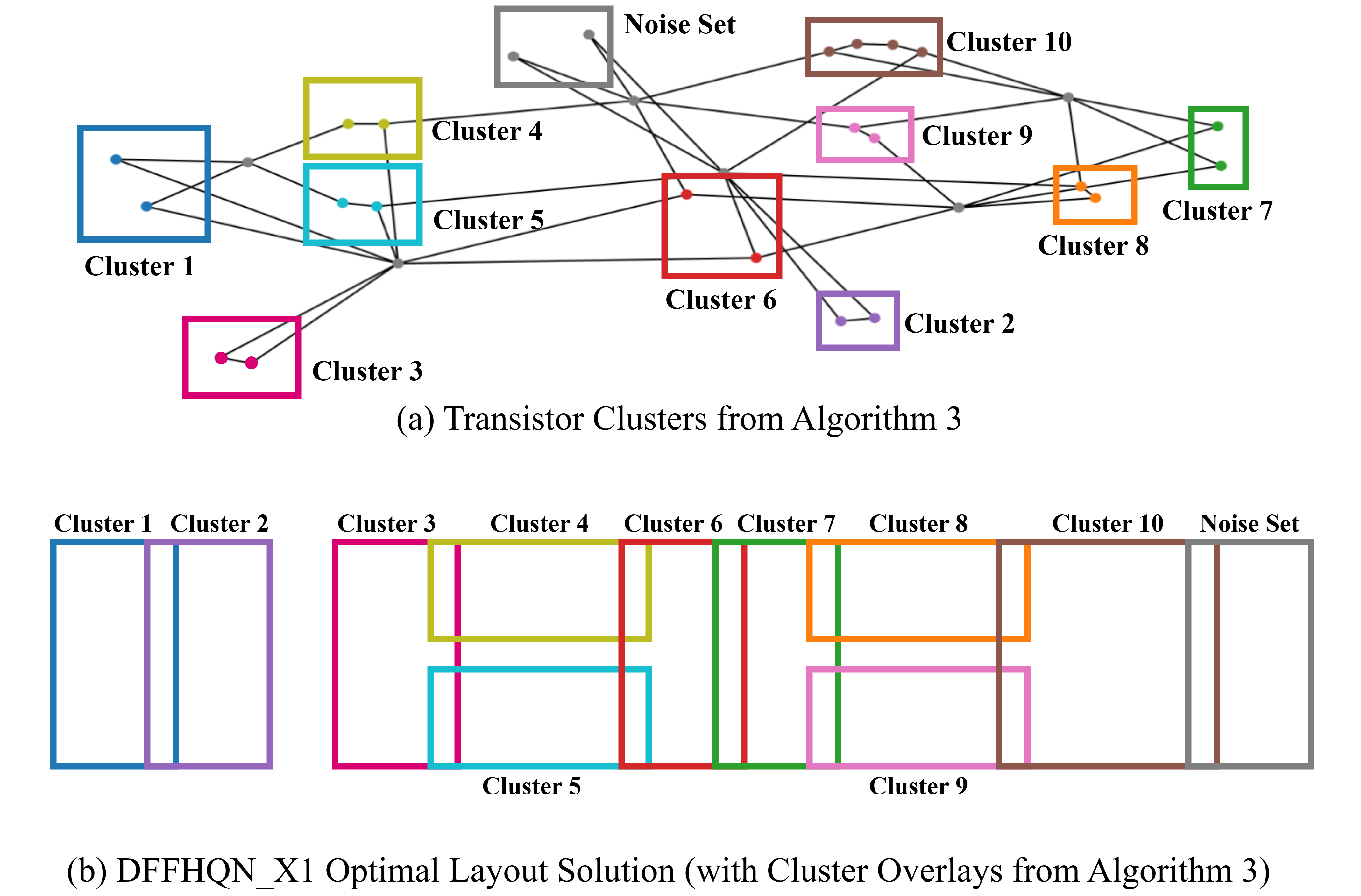}
  \caption{Algorithm~\ref{alg:kkhdb} reveals transistor clusters found 
  in the optimal layout solution for DFFHQN\_X1. (a) Identified 10 transistor 
  clusters and one noise set via Algorithm~\ref{alg:kkhdb}. (b) The optimal 
  layout solution generated without cluster-based constraints. The colored 
  overlays show the correspondence between the clusters 
  identified by Algorithm~\ref{alg:kkhdb} and the clusters in the optimal placement.}
  \label{fig:cluster}
\end{figure}

We cluster transistors whose diffusion 
should be shared and impose structural constraints. 
% Prior work uses machine learning 
% techniques~\cite{ho2024transformerclustering} or Euler-path 
% enumeration~\cite{Guo2025ISEDA} to discover transistor clusters, often introducing 
% substantial algorithmic complexity or external dependencies (i.e., training data). 
Our approach is designed as a lightweight add-on to constraint solving.
Figure~\ref{fig:cluster} shows a one-to-one correspondence between the clusters 
from our method and the clusters in the optimal ground-truth layout of DFFHQN\_X1. By 
utilizing these clusters to constrain the search space, the solver 
operates at the cluster-level rather than on individual 
transistors. Our clustering procedures are detailed in 
Algorithm~\ref{alg:kkhdb} and summarized below:

\begin{algorithm}[tb]
\caption{Graph Construction and Clustering}
\label{alg:kkhdb}
\Input{Netlist, cluster bounds $[k_{\min}, k_{\max}]$}
\Output{set of transistor clusters $Clst$}

$M \gets \textsc{BuildCircuitGraph}(\text{Netlist})$\;
$M \gets M \setminus \{\text{VDD}, \text{VSS}\}$

\ForEach{net node $n \in M$ such that $\deg(n) \le 2$}{
    $\{tr_i, tr_j\} \gets \textsc{Neighbors}(M, n)$\;
    $\textsc{AddEdge}(M, tr_i, tr_j)$\;
    $M \gets M \setminus \{n\}$\;
}

\For{net node $n \in M$}{
    \For{neighbor $tr \in \textsc{Neighbors}(M, n)$}{
        $w(n, tr) \gets \deg(n)$
    }
}

$\mathbb{X} \gets \textsc{KamadaKawaiLayout}(M, w)$

$Clst, noise \gets \textsc{HDBSCAN}(M, \mathbb{X}, k_{\min}, k_{\max})$\;
\Return $Clst$\;
\end{algorithm}

% \new{\noindent\textbf{Step 1}:
\noindent\emph{Lines 1--8:} We build a graph \(M\) with both transistors and nets 
represented as vertices. Connections are established exclusively between transistor 
vertices and net vertices: a link exists between a transistor \(tr\) and a net \(n\) 
if the net is incident to the transistor's source, drain, or gate. We remove 
power and ground net nodes. We then remove any net node with degree \(\le 2\), and reconnect 
the incident transistors directly. This representation directly correlates with 
the potential candidate clusters for diffusion sharing and placement proximity.

\noindent\emph{Lines 9--12:} We assign weights to fan-out edges from
each net node proportional to the net degree and compute a two-dimensional
embedding using the Kamada-Kawai spring model~\cite{kamada_kawai}. The model
optimizes a spring-based energy function to preserve graph distances in
Euclidean space, producing a layout in which strongly connected transistors
are placed in close proximity.

\noindent\emph{Line 13--14:} We apply Hierarchical Density-Based 
Spatial Clustering of Applications with Noise ~~(HDBSCAN)~\cite{hdbscan}\cite{scikit-learn} 
to the embedded nodes, using the pairwise Euclidean distances in the 
embedding space as the clustering metric. HDBSCAN is chosen because 
it identifies density-based clusters without requiring a predefined 
cluster count, and explicitly models a noise set. Transistors classified 
as \textit{noise} are isolated or weakly connected in the embedding 
space and are treated as free-floating to preserve satisfiability. 
HDBSCAN also accepts minimum 
and maximum cluster sizes, denoted as $k_{min}$ and $k_{max}$. $k_{min}$ 
is always set to two as it is the minimum effective cluster size. $k_{max}$ 
can range from two to six, depending on the netlist. When $k_{max}=6$, the 
search space is aggressively pruned, but the solver may produce suboptimal 
solutions or even become unsatisfiable. Setting $k_{max} = 2$ is 
safer but produces limited speedup. The effect of these parameters 
is studied in Section~\ref{subsec:accstudy}. 
HDBSCAN returns a noise set of transistors and a set of density-based 
clusters \(\{Clst_1, Clst_2, \ldots\}\), where each cluster \(Clst_k\) 
captures a group of transistors with strongly coupled connectivity. 
The noise set is discarded as it may over-constrain the layout 
and result in an unsatisfiable search space. \ywrone{Because clustering 
restricts the encoded feasible set, it does not always guarantee 
optimality with respect to the original formulation.}

% \footnote{\yw{Transistors in the noise set are either isolated or loosely connected in the embedding space. We treat them as free-floating to preserve solver flexibility.}}

Once transistor clusters are identified, we enforce diffusion 
continuity within each cluster. Let $Clst_k$ be a cluster comprising a set of 
transistors \(\mathcal{T}_k\). For each transistor \({tr} \in \mathcal{T}_k\), let \(x_{tr}\) 
denote its left coordinate and \(w_{tr}\) its diffusion width. 
We enforce contiguous placement by constraining the total span of the cluster 
to equal the sum of the individual transistor widths:
\begin{equation}\label{eq:cluster}
\left( \max_{{tr} \in \mathcal{T}_k} (x_{tr} + w_{tr}) \right) - \min_{{tr} \in \mathcal{T}_k} x_{tr} 
\;=\; 
\sum_{{tr} \in \mathcal{T}_k} w_{tr}.
\end{equation}
The term \(\max(x_{tr} + w_{tr}) - \min(x_{tr})\) represents the width of the
bounding box of $Clst_k$. Enforcing equality with the sum of widths
eliminates spacing gaps and overlaps, ensuring a single
uninterrupted diffusion chain. \ywrone{Each PMOS-only and NMOS-only cluster constraint 
is enforced by Equation~\ref{eq:cluster}. For a cluster of size $|\mathcal{T}_k| = 2$ 
containing one PMOS and one NMOS transistor, we additionally require gate sharing ($scx_c = 0$).}
For clusters \ywrone{of size $|\mathcal{T}_k| > 2$}, containing both PMOS and NMOS
devices, the total width \(\sum w_{tr}\) is determined by the transistor type
with the maximum aggregate diffusion width; transistors of the complementary type
are constrained to lie within this spatial bound.

\subsection{Identical Transistor Partitioning (ITP)} \label{subsec:partition}

High-drive-strength cells (e.g., X2, X4 and X8 variants) contain groups 
of electrically identical transistors to achieve the target drive strength. 
Without additional partitioning, the solver must explore every 
permutation of these identical transistors, even though many such permutations 
yield identical layouts. CPCell partitions 
identical transistors into a canonical ordering. 
Given a set $S$ of transistors with identical gate, source, drain and 
type attributes, we enforce ordering constraints of the form:
\begin{equation}
x_{{tr}_1} \le x_{{tr}_2} \le \cdots \le x_{tr_{|S|}},
\end{equation}
where $x_{tr}$ denotes the placement coordinate. 
This symmetry breaking explores only one layout per permutation class.

\begin{figure}[ht]
\centering
\includegraphics[width=0.9\columnwidth]{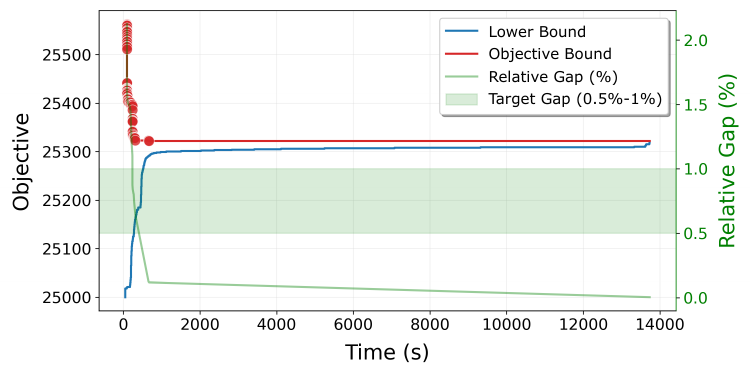}
\caption{Search progress of the CP-SAT solver for DFFHQN\_X1 with 
cell-width ($CW$) and wirelength ($WL$) objectives. The red line graph denotes the 
objective, where each marker corresponds to the discovery of 
a new feasible layout, while the blue line graph represents the lower bound.}
\label{fig:search}
\end{figure}

\subsection{Routing Lower Bound Tightening (RLBT)} \label{subsec:lowerbound}

Half-perimeter wirelength (HPWL)~\cite{timberwolf} from intermediate
feasible placements provides a routing lower bound for faster optimality
verification. CP-SAT combines branch-and-bound, linear programming (LP)
relaxations and Boolean satisfiability (SAT) reasoning~\cite{perron_cpsatlp}.
In Figure~\ref{fig:search}, red denotes the best feasible objective and
blue the relaxation-derived lower bound. Branching, constraint propagation
and relaxation refinement tighten the bounds; convergence certifies optimality.

Empirically, we observe that CP-SAT often identifies an optimal or near-optimal solution early 
in the search, whereas a substantial portion of the remaining runtime is devoted 
to tightening the lower bound to certify optimality. If the lower bound can be explicitly 
derived early on, we can reduce the runtime for such a certification. At block-level, 
HPWL is commonly used as a proxy for placement quality. 
At cell-level, we can use HPWL to estimate the wirelength 
of each net from the intermediate feasible placement solution, which serves as a routing lower bound. 
Once CP-SAT converges on a placement, this estimation immediately strengthens the lower bound, thereby 
accelerating optimality verification. 

\subsection{Early Termination \& Relative Gap} \label{subsec:earlyterm}

To reduce solver runtime for large problem instances, we adopt an early-termination 
criterion based on the \emph{relative gap} between the current objective 
bound and the lower bound. In our experiments, 
the relative gap is set to $0.5\%$ or $1\%$ for large netlists and is defined as
\begin{equation}
\text{Relative Gap} = \frac{|\text{ObjectiveBound} - \text{LowerBound}|}{|\text{ObjectiveBound}|}.
\end{equation}

From Figure~\ref{fig:search}, we observe that after the initial rapid objective 
improvement before 672\,s, no better solution was found; instead, the 
solver devotes $13,062$\,s exclusively to tightening the lower bound to 
formally certify optimality. This behavior motivates the use of the relative gap for early termination.
When the relative gap is used, we only guarantee optimality for the primary objective, cell width.

% \new{To preserve the optimality of the primary objective, cell width $CW$, we require the 
% absolute optimality gap to remain below $\frac{\lambda_0}{2} = 500$. For netlists with more than 
% 40 transistors, where $CW \ge 40$, the total objective value is at least 
% $40,000$ (given $\lambda_0 = 1000$). In these cases, a $1\%$ relative gap translates to an 
% absolute gap of $400$, which is strictly less than the $500$ threshold. This approach 
% accelerates convergence while ensuring that early termination never compromises the optimal cell width.}

\section{Cell-level Experiments on Gear-ratio-enabled Cell Layouts}\label{sec:cell-exp}
\ywrone{All cells are generated under the predictive 3-nm-class technology
defined by the PROBE3.0 framework~\cite{choi2023probe30}; Table~\ref{tab:techparams}
summarizes the process-technology parameters.}
\begin{table}[t]
\centering
\caption{\ywrone{Process technology parameters from PROBE3.0~\cite{choi2023probe30}.}}
\label{tab:techparams}
\ywrone{
\begin{tabularx}{\columnwidth}{|p{4.0cm}|X|}
\hline
\textbf{Parameter} & \textbf{Value} \\ \hline
Contacted poly pitch ($CPP$) & 45\,nm \\ \hline
M0 pitch ($M0P$) & 24\,nm \\ \hline
M1 pitch ($M1P$) & 45\,nm, 30\,nm, 27\,nm \\ \hline
M2 pitch ($M2P$) & 24\,nm \\ \hline
Fin count & 2 \\ \hline
Cell height & 144\,nm (4 routing tracks) \\ \hline
BEOL stack & 14 layers (M0--M13) \\ \hline
Design rules & Min. EOL=21\,nm, MAR=595\,nm$^2$ \\ \hline
Supply volt. / corner / temperature & 0.7\,V, $TT$, $25^{\circ}$C \\ \hline
\end{tabularx}
}
\end{table}
Cell-level experiments were conducted on an Intel Core
i9-12900K CPU with 64 GB RAM using CP-SAT solver (v9.14)~\cite{ortools} 
with six parallel search workers. All experiments target 
the 2-Fin 4-Routing Track (2F4T)
configuration, whose horizontal routing is more constrained than
in the previously studied 3-Fin 5-Routing Track (3F5T)
configuration~\cite{agr}, better exposing gear-ratio effects.
All layouts use the single-diffusion-break setting, reducing
transistor spacing and cell width. We analyze representative cells
across gear ratios and offsets, ablate objective functions and
weights using DFFHQN\_X1, and evaluate acceleration techniques
on several D flip-flops and a high-drive-strength buffer.

\subsection{Cell Metrics}\label{subsec:maincell}

We present a sensitivity study on gear ratios in 
Table~\ref{tab:main_results_expanded} \ywrone{and an averaged full library 
comparison with previous gear-ratio-enabled work in Table~\ref{tab:complete_results}}. 
In Table~\ref{tab:main_results_expanded}, we 
report results for 22 \ywrone{high-drive-strength and functionally diverse} cells out of the 
40 cells~\cite{agr} designed under 2F4T configuration. We evaluate 
three gear ratio (GR) settings--45:45 (1:1), 45:30 (3:2) 
and 45:27 (5:3)--along with their respective offset counterparts. 
For a given GR parameter, the number of valid offset 
choices is determined by the alignment between the poly grid 
(pitch $mp^1$) and the M1 grid (pitch $mp^3$). Specifically, 
admissible offsets are restricted to integer multiples of the greatest 
common divisor (gcd), or
$\gcd(mp^3, mp^1)$ within one M1 period. 
% Any shift that violates 
% this condition causes M1 grid points to fall between poly grid 
% points, resulting in a misaligned column that cannot be 
% legally placed. 
% Conversely, offsets of the form 
% $k \cdot \gcd(mp^3, mp^1)$ preserve periodic alignment and thus 
% constitute all legal offset candidates. 
1:1 GR admits no valid offset, whereas 3:2 GR has 
one valid offset of 15\,nm (3:$2^{\dagger}$), and 5:3 GR has 
two valid offsets of 9\,nm and 18\,nm (5:$3^{\dagger}$ and 5:$3^{\dagger\dagger}$).

Across all settings, 1:1 GR consistently achieves 
the shortest CP-SAT runtime due to its smaller routing search space. 
However, its limited routing resources result in the largest average cell width. 
Across the 22 reported cells, non-1:1 GR configurations reduce 
average cell width from 7.82 CPP to 7.68–7.73 CPP. The two 5:3 GR offset 
configurations have the lowest average width, including a one-CPP reduction 
for MUX2\_X1. The mean per-cell normalized wirelength is slightly higher 
than the 1:1 baseline, and the effect of GR and offset vary by cell. The 
1:1 GR configuration has the lowest average runtime in this experiment.

\ywrone{In Table~\ref{tab:complete_results}, we compare our framework 
against PROBE3.0~\cite{choi2023probe30} and SMTCell-GR~\cite{agr} across the full set of 40 cells. 
Our approach achieves smaller average cell widths and normalized wirelengths, 
comparable or lower average M2 usage, and up to a \(9.47\times\) 
runtime speedup over the available baselines, while 
additionally supporting offset gear-ratio configurations.}

% For certain logic cells such as AOI21\_X1 and OAI21\_X1, tighter GR 
% and their offset variants enable additional routing flexibility that 
% allows the solver to realize more compact placements and achieve 
% reduced cell width. Taken together, these results indicate that the 
% 45:30 (3:2) GR, with optional offset enablement, provides the most 
% favorable balance between wirelength reduction, width compaction, 
% and computational complexity among the evaluated parameters.

\begin{table*}[hbt!]
\caption{Cell metrics and runtimes for 2F4T gear-ratio-enabled cells: 1:1 GR has no offset variant; 3:2 GR has 
one 15\,nm offset (3:$2^{\dagger}$); and 5:3 has two offsets 
of 9\,nm and 18\,nm (5:$3^{\dagger}$ and 5:$3^{\dagger\dagger}$), respectively. Wirelength is normalized to the 1:1 GR baseline.}
\label{tab:main_results_expanded}
\tabcolsep = 3pt % Reduced slightly to fit more columns
\centering
\resizebox{1.0\textwidth}{!}{%
\begin{tabular}{|lrr|*{6}{r}|*{6}{r}|*{6}{r}|*{6}{r}|}
\hline
\multirow{2}{*}{\textbf{Cell Name}} &
\multirow{2}{*}{\textbf{\#FET}} &
\multirow{2}{*}{\textbf{\#Net}} &
\multicolumn{6}{c|}{\textbf{Cell Width (CPP)}} &
\multicolumn{6}{c|}{\textbf{Normalized Wirelength}} &
\multicolumn{6}{c|}{\textbf{\#M2}} &
\multicolumn{6}{c|}{\textbf{CP-SAT Runtime (s)}} \\
 & & &
1:1 & 3:2 & 3:$2^\dagger$ & 5:3 & 5:$3^\dagger$ & 5:3$^{\dagger\dagger}$ &
1:1 & 3:2 & 3:$2^\dagger$ & 5:3 & 5:$3^\dagger$ & 5:3$^{\dagger\dagger}$ &
1:1 & 3:2 & 3:$2^\dagger$ & 5:3 & 5:$3^\dagger$ & 5:3$^{\dagger\dagger}$ &
1:1 & 3:2 & 3:$2^\dagger$ & 5:3 & 5:$3^\dagger$ & 5:3$^{\dagger\dagger}$ \\
\hline

AND2\_X2   & 8  & 7   
& 5 & 5 & 5 & 5 & 5 & 5   
% & 466.0 & 412.0 & 421.0 & 412.0 & 412.0 & 421.0
& 1.00 & 0.88 & 0.90 & 0.88 & 0.88 & 0.90
& 0 & 0 & 0 & 0 & 0 & 0     
& 2 & 5 & 4 & 5 & 4 & 4 \\
AND3\_X2   & 10 & 9   
& 6 & 6 & 6 & 6 & 6 & 6 
% & 392.5 & 374.5 & 385.0 & 374.5 & 383.5 & 388.0 
& 1.00 & 0.95 & 0.98 & 0.95 & 0.98 & 0.99
& 0 & 0 & 0 & 0 & 0 & 0 
& 5 & 7 & 6 & 7 & 7 & 8 \\
% AOI21\_X1  & 6  & 8   
% & 5 & 4 & 4 & 4 & 4 & 4 
% % & 531.0 & 409.5 & 426.0 & 409.5 & 423.0 & 409.5 
% & 1.00 & 0.77 & 0.80 & 0.77 & 0.80 & 0.77
% & 0 & 0 & 0 & 0 & 0 & 0 
% & 2 & 2 & 2 & 3 & 3 & 3 \\
\ywrone{AOI21\_X2} & 12  &  8   
& 8 & 7  & 7  & 7  & 7 & 7
& 1.00 & 1.53 & 1.64  & 1.36 & 1.37 & 1.40
& 0 & 2 & 2 & 1 & 1 & 1
& 15 & 22 & 32 & 27 & 26 & 42 \\
AOI22\_X2  & 16 & 10  
& 10 & 10 & 10 & 10 & 10 & 10 
% & 1,401.0 & 1,401.0 & 1,401.0 & 1,401.0 & 1,548.0 & 1,401.0  
& 1.00 & 1.00 & 1.00 & 1.00 & 1.10 & 1.00
& 0 & 0 & 0 & 0 & 1 & 0 
& 53 & 533 & 210 & 278 & 136 & 252 \\
BUF\_X4    & 10 & 5   
& 7 & 7 & 7 & 7 & 7 & 7 
% & 397.5 & 397.5 & 397.5 & 397.5 & 397.5 & 397.5 
& 1.00 & 1.00 & 1.00 & 1.00 & 1.00 & 1.00
& 0 & 0 & 0 & 0 & 0 & 0 
& 8 & 18 & 14 & 18 & 29 & 20 \\
BUF\_X8    & 20 & 5   
& 13 & 13 & 13 & 13 & 13 & 13 
% & 937.5 & 937.5 & 937.5 & 937.5 & 937.5 & 937.5 
& 1.00 & 1.00 & 1.00 & 1.00 & 1.00 & 1.00
& 0 & 0 & 0 & 0 & 0 & 0 
& 210 & 712 & 613 & 1105 & 2931 & 991 \\
DFFHQN\_X1 & 24 & 17  
& 14 & 14 & 14 & 14 & 14 & 14 
% & 2,587.5 & 2,575.5 & 2,631.5 & 2,575.5 & 2,589.0 & 2,611.5 
& 1.00 & 1.00 & 1.02 & 1.00 & 1.00 & 1.01
& 2 & 2 & 2 & 2 & 2 & 2 
& 121 & 254 & 300 & 352 & 429 & 414 \\
INV\_X4    & 8  & 4   
& 5 & 5 & 5 & 5 & 5 & 5 
% & 265.0 & 265.0 & 265.0 & 265.0 & 265.0 & 265.0 
& 1.00 & 1.00 & 1.00 & 1.00 & 1.00 & 1.00
& 0 & 0 & 0 & 0 & 0 & 0 
& 1 & 1 & 2 & 1 & 3 & 3 \\
INV\_X8    & 16 & 4   
& 9 & 9 & 9 & 9 & 9 & 9 
% & 625.0 & 625.0 & 625.0 & 625.0 & 625.0 & 625.0 
& 1.00 & 1.00 & 1.00 & 1.00 & 1.00 & 1.00
& 0 & 0 & 0 & 0 & 0 & 0 
& 1 & 1 & 2 & 3 & 3 & 3 \\
LHQ\_X1    & 8  & 13  
& 10 & 10 & 10 & 10 & 10 & 10 
% & 1,614.5 & 1,616.0 & 1,639.5 & 1,616.0 & 1,580.0 & 1,616.0 
& 1.00 & 1.00 & 1.02 & 1.00 & 0.98 & 1.00
& 1 & 1 & 1 & 1 & 1 & 1 
& 28 & 141 & 256 & 111 & 353 & 250 \\
MUX2\_X1   & 12 & 12  
& 8 & 8 & 8 & 8 & 7 & 7 
% & 1,257.0 & 1,168.5 & 1,270.5 & 1,168.5 & 1,123.5 & 1,177.5 
& 1.00 & 0.93 & 1.01 & 0.93 & 0.89 & 0.94
& 1 & 1 & 1 & 1 & 1 & 1 
& 9 & 9 & 13 & 26 & 9 & 9 \\
NAND2\_X2  & 8  & 6   
& 5 & 5 & 5 & 5 & 5 & 5 
% & 440.0 & 440.0 & 440.0 & 440.0 & 440.0 & 440.0 
& 1.00 & 1.00 & 1.00 & 1.00 & 1.00 & 1.00
& 0 & 0 & 0 & 0 & 0 & 0 
& 3 & 4 & 5 & 6 & 6 & 6 \\
NAND3\_X2  & 12 & 7   
& 7 & 7 & 7 & 7 & 7 & 7 
% & 1,166.0 & 1,173.5 & 1,173.5 & 1,170.5 & 1,170.5 & 1,170.5 
& 1.00 & 1.00 & 1.00 & 1.00 & 1.00 & 1.00
& 1 & 1 & 1 & 1 & 1 & 1 
& 12 & 24 & 28 & 45 & 51 & 50 \\
NAND4\_X2  & 16 & 9   
& 9 & 9 & 9 & 9 & 9 & 9 
% & 1,794.5 & 1,809.5 & 1,809.5 & 1,821.5 & 1,794.5 & 1,794.5 
& 1.00 & 1.00 & 1.00 & 1.00 & 1.00 & 1.00
& 1 & 1 & 1 & 1 & 1 & 1 
& 15 & 53 & 47 & 115 & 71 & 66 \\
NOR2\_X2   & 12 & 8   
& 5 & 5 & 5 & 5 & 5 & 5 
% & 440.0 & 440.0 & 440.0 & 440.0 & 440.0 & 440.0
& 1.00 & 1.00 & 1.00 & 1.00 & 1.00 & 1.00
& 0 & 0 & 0 & 0 & 0 & 0 
& 3 & 5 & 5 & 6 & 6 & 6 \\
NOR3\_X2   & 12 & 7   
& 7 & 7 & 7 & 7 & 7 & 7 
% & 1,166.0 & 1,173.5 & 1,173.5 & 1,170.5 & 1,170.5 & 1,170.5
& 1.00 & 1.00 & 1.00 & 1.00 & 1.00 & 1.00
& 1 & 1 & 1 & 1 & 1 & 1 
& 11 & 23 & 35 & 33 & 56 & 41 \\
NOR4\_X2   & 16 & 9   
& 9 & 9 & 9 & 9 & 9 & 9 
% & 1,794.5 & 1,809.5 & 1,809.5 & 1,821.5 & 1,794.5 & 1,794.5 
& 1.00 & 1.00 & 1.00 & 1.00 & 1.00 & 1.00
& 1 & 1 & 1 & 1 & 1 & 1 
& 19 & 61 & 58 & 122 & 97 & 123 \\
% OAI21\_X1  & 6  & 8   
% & 5 & 4 & 4 & 4 & 4 & 4 
% % & 531.0 & 409.5 & 426.0 & 409.5 & 423.0 & 409.5 
% & 1.00 & 0.77 & 0.80 & 0.77 & 0.80 & 0.77
% & 0 & 0 & 0 & 0 & 0 & 0 
% & 2 & 2 & 2 & 3 & 3 & 3 \\
\ywrone{OAI21\_X2} & 12  &  8
& 8 & 7  & 7  & 7  & 7 & 7
& 1.00 & 1.53 & 1.64  & 1.36 & 1.37 & 1.40
& 0 & 2 & 2 & 1 & 1 & 1 
& 15 & 36 & 17 & 29 & 45 & 21 \\
OAI22\_X2  & 16 & 10  
& 10 & 10 & 10 & 10 & 10 & 10 
% & 1,401.0 & 1,401.0 & 1,401.0 & 1,401.0 & 1,548.0 & 1,401.0 
& 1.00 & 1.00 & 1.00 & 1.00 & 1.10 & 1.00
& 0 & 0 & 0 & 0 & 1 & 0 
& 48 & 456 & 167 & 297 & 73 & 266 \\
OR2\_X2    & 8  & 7   
& 5 & 5 & 5 & 5 & 5 & 5 
% & 488.5 & 413.5 & 421.0 & 412.5 & 407.5 & 425.5 
& 1.00 & 0.85 & 0.86 & 0.84 & 0.83 & 0.87
& 0 & 0 & 0 & 0 & 0 & 0 
& 3 & 5 & 4 & 5 & 5 & 4 \\
OR3\_X2    & 10 & 9   
& 6 & 6 & 6 & 6 & 6 & 6 
% & 392.5 & 377.5 & 385.0 & 374.5 & 383.5 & 388.0
& 1.00 & 0.96 & 0.98 & 0.95 & 0.98 & 0.99
& 0 & 0 & 0 & 0 & 0 & 0 
& 5 & 9 & 7 & 11 & 8 & 8 \\
XOR2\_X1   & 10 & 9   
& 6 & 6 & 6 & 6 & 6 & 6 
% & 1,153.0 & 1,138.0 & 1,153.0 & 1,135.0 & 1,130.5 & 1,139.5 
& 1.00 & 0.99 & 1.00 & 0.98 & 0.98 & 0.99
& 1 & 1 & 1 & 1 & 1 & 1 
& 4 & 6 & 6 & 9 & 7 & 6 \\

\hdashline
\textbf{Average} & & &
7.82 & 7.73 & 7.73 & 7.73 & 7.68 & 7.68 &
% 940.98 & 918.82 & 931.77 & 919.32 & 928.77 & 922.16 &
1.00 & 1.03 & 1.05 & 1.01 & 1.02 & 1.02 &
0.41 & 0.59 & 0.59 & 0.50 & 0.59 & 0.50 &
26.86 & 108.41 & 83.23 & 118.68 & 197.95 & 117.86 \\
\hline
\end{tabular}
}
\end{table*}

\begin{table*}[hbt!]
\ywrone{
\caption{\ywrone{Comparison of averaged cell metrics and total
runtimes across PROBE3.0, SMTCell-GR, and our work for 
2F4T gear-ratio (GR) enabled cells. The 1:1 GR has no offset variant; 
the 3:2 GR includes a 15\,nm offset (3:$2^{\dagger}$); and the 
5:3 GR includes 9\,nm (5:$3^{\dagger}$) and 18\,nm (5:$3^{\dagger\dagger}$) 
offsets. Wirelength is normalized to the 1:1 GR baseline. 
``---'': unavailable configuration for the tool.}}
\label{tab:complete_results}
\tabcolsep = 3pt % Reduced slightly to fit more columns
\centering
\resizebox{1.0\textwidth}{!}{%
\begin{tabular}{|l|*{6}{r}|*{6}{r}|*{6}{r}|*{6}{r}|}
\hline
\multirow{2}{*}{\textbf{Tool}} &
\multicolumn{6}{c|}{\textbf{Average Cell Width (CPP)}} &
\multicolumn{6}{c|}{\textbf{Average Normalized Wirelength}} &
\multicolumn{6}{c|}{\textbf{Average \#M2}} &
\multicolumn{6}{c|}{\textbf{Total Runtime (s)}} \\
 &
1:1 & 3:2 & 3:$2^\dagger$ & 5:3 & 5:$3^\dagger$ & 5:3$^{\dagger\dagger}$ &
1:1 & 3:2 & 3:$2^\dagger$ & 5:3 & 5:$3^\dagger$ & 5:3$^{\dagger\dagger}$ &
1:1 & 3:2 & 3:$2^\dagger$ & 5:3 & 5:$3^\dagger$ & 5:3$^{\dagger\dagger}$ &
1:1 & 3:2 & 3:$2^\dagger$ & 5:3 & 5:$3^\dagger$ & 5:3$^{\dagger\dagger}$ \\
\hline
PROBE3.0~\cite{choi2023probe30}  
&--- & 6.60     & --- & --- & --- & --- 
% &--- & 1027.63  & --- & --- & --- & --- 
&--- & 1.16  & --- & --- & --- & --- 
&--- & 0.35& --- & --- & --- & ---  
&--- & 20518.00& --- & --- & --- & ---  \\ 
SMTCell-GR~\cite{agr}
&  6.83 &6.50 & ---& 6.40 & --- & --- 
% & 1131.18&1155.93 & --- & 1154.13 & --- & --- 
& 1.64&1.31 & --- & 1.40 & --- & --- 
& 0.35& 0.38 & --- & 0.38 & --- & ---  
& 7136.49 & 12946.25 & --- &  11481.80& --- & ---  \\ 
Our work
& 6.43& 6.08& 6.56& 6.08& 6.56& 6.56
% & 688.01& 884.53& 1240.28& 823.68& 1240.34& 1203.00 
& 1.00& 1.00& 1.00& 1.00& 1.00& 1.00 
& 0.23& 0.35 & 0.75& 0.23& 0.81& 0.63  
& 753.30 & 2477.40 & 2584.32 & 2821.10 & 4952.41 & 3762.09 \\ 
 \hline
\end{tabular}
}
}
\end{table*}

\subsection{Ablation Study on Objectives}~\label{subsec:ablation}
We ablate auxiliary objectives on DFFHQN\_X1 to evaluate their
individual and combined effects on solver runtime. 
We perform comparison studies of five objective combinations 
to isolate the individual and combined effects of auxiliary 
objectives on solver convergence behavior:
\begin{enumerate}[label=(\alph*)]
  \item $\lambda_0 \cdot CW + \lambda_1 \cdot WL$  \label{ala:base}
  \item $\lambda_0 \cdot CW + \lambda_1 \cdot WL - \lambda_2 \cdot SGD$  \label{ala:SGD}
  \item $\lambda_0 \cdot CW + \lambda_1 \cdot WL - \lambda_3 \cdot DBX$
  \label{ala:DBX}
  \item $\lambda_0 \cdot CW + \lambda_1 \cdot WL + \lambda_4 \cdot M2$
  \label{ala:M2}
  \item $\lambda_0 \cdot CW + \lambda_1 \cdot WL - \lambda_2 \cdot SGD - \lambda_3 \cdot DBX+ \lambda_4 \cdot M2$  \label{ala:all}
\end{enumerate}

Study~\ref{ala:base} considers only cell width and wirelength 
minimization and serves as the baseline configuration. 
Studies~\ref{ala:SGD}, \ref{ala:DBX} and \ref{ala:M2} individually 
activate source/gate/drain sharing, diffusion break columns 
and M2 usage, respectively, to isolate 
their effects on solver convergence. Study~\ref{ala:all} enables 
all auxiliary objectives to evaluate 
their combined impact on accelerating solver convergence.

\begin{figure}[ht]
\centering
\includegraphics[width=0.9\columnwidth]{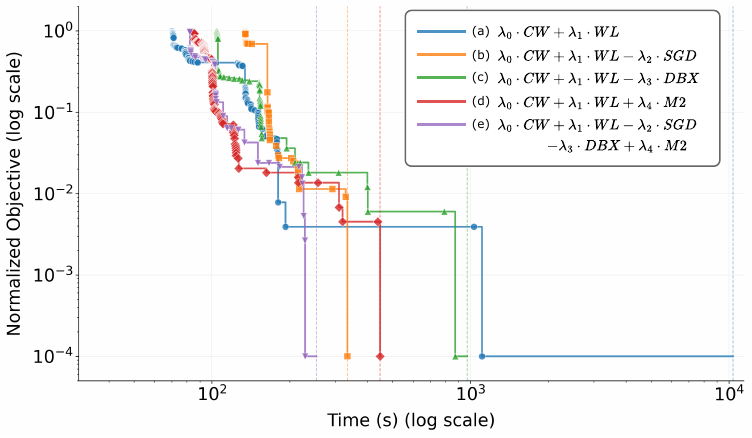}
\caption{Search progress of the CP-SAT solver for DFFHQN\_X1 under different 
objective sets, shown on a logarithmic time scale with normalized objective 
values. The vertical dashed line marks the end of the search.}
\label{fig:batchobj}
\end{figure}

% \new{Figure~\ref{fig:batchobj} compares CP-SAT convergence 
% under different objective configurations for DFFHQN\_X1. 
% All configurations converge to the same final cell width 
% and wirelength, confirming that auxiliary objectives do 
% not affect layout optimality. However, convergence speed 
% differs significantly. Enabling individual acceleration 
% objectives in Studies~\ref{ala:SGD}, \ref{ala:DBX}, and 
% \ref{ala:M2} substantially reduces runtime relative to 
% the baseline. Enabling all auxiliary objectives jointly 
% in Study~\ref{ala:all} yields the fastest convergence 
% overall. These results demonstrate that carefully 
% designed auxiliary objectives improve solver efficiency 
% without degrading layout quality.}

Figure~\ref{fig:batchobj} compares CP-SAT convergence for DFFHQN\_X1 
under different objective configurations.
All configurations converge to the same final cell width and wirelength, confirming that 
auxiliary objectives do not affect layout optimality. However, convergence 
speed differs significantly. Enabling individual acceleration objectives 
substantially reduces runtime relative to the baseline: 
Study~\ref{ala:SGD} achieves a 96.7\% speedup, Study~\ref{ala:DBX} achieves 
a 90.6\% speedup and Study~\ref{ala:M2} achieves a 95.7\% speedup. 
Finally, combining all auxiliary objectives in 
Study~\ref{ala:all} yields the fastest convergence overall, 
achieving a 97.6\% speedup. These results demonstrate that 
carefully designed auxiliary objectives can improve solver 
efficiency without degrading layout quality.

\subsection{Effect of M0 Pin Accessibility Constraints}~\label{subsec:m0pinstudy}
\begin{figure}[ht]
\centering
\includegraphics[width=0.85\columnwidth]{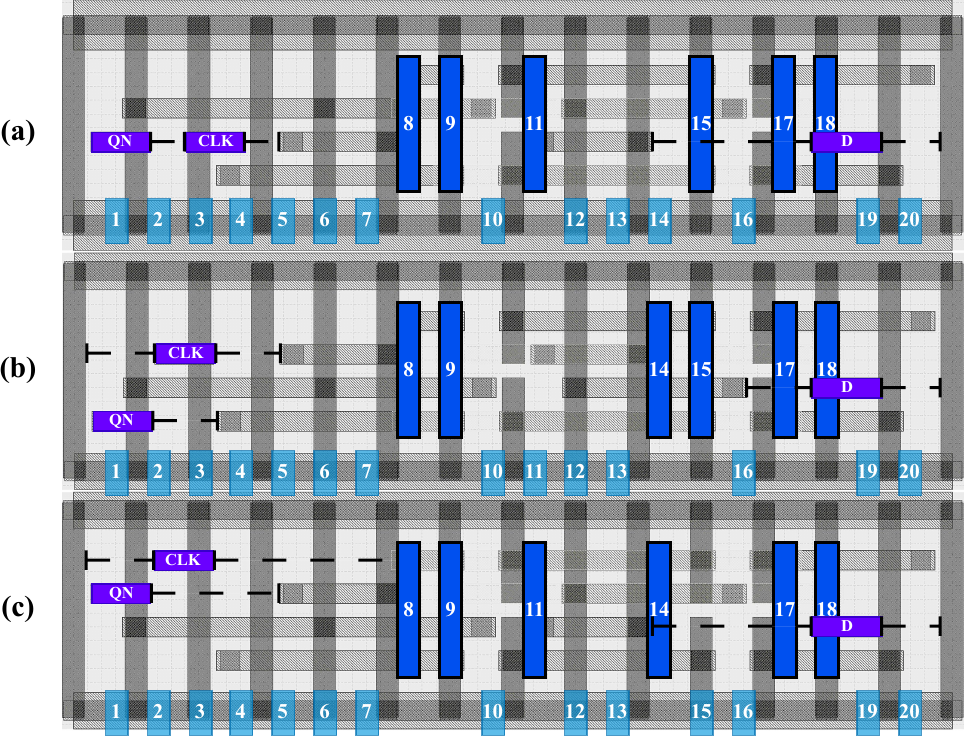}
\caption{Comparison of pin accessibility for DFFHQN\_X1: 
(a) baseline, (b) PS enabled and (c) PS and MPO ($\Theta$ = 4) enabled. 
M0 pins are shown in purple with dashed extensible regions; 
black-outlined M1 tracks indicate routing blockages and 
light-blue segments denote candidate access locations.}
\label{fig:pinaccess}
\end{figure}
We model M0 pin accessibility using the count of unblocked 
on-grid access points, approximated by the number of M1 routing tracks 
that overlap each extended M0 pin. Prior works~\cite{KahngKLX22} \cite{SunCCHC18} show that alignment 
to routing tracks and the resulting number of on-grid access 
points strongly correlate with pin accessibility, so the overlap count 
serves as a simple proxy. 
Figure~\ref{fig:pinaccess} compares the DFFHQN\_X1 layouts under 
progressively enabled pin accessibility constraints. 
We measure M0 pin accessibility by the number of unblocked M1 routing tracks 
that geometrically overlap each extended M0 pin. 
Table~\ref{tab:m0_pin_access} details the specific track counts 
and indices for each configuration. 
Figure~\ref{fig:pinaccess}(a) shows the baseline where 
pins share a single row, limiting track access. 
Figure~\ref{fig:pinaccess}(b) shows the effect of PS, which distributes pins across 
three rows and reduces interference. Figure~\ref{fig:pinaccess}(c) shows the highest 
accessibility achievable by combining PS with MPO ($\Theta$ = 4). M0 pins extend to 
a minimum of four M1 access points, with the CLK pin reaching seven points.

\begin{table}[ht] 
\centering 
\caption{Number and indices of M1 routing tracks overlapping each 
extended M0 pin for DFFHQN\_X1 under PS and MPO ($\Theta$).} \label{tab:m0_pin_access} 
\resizebox{0.8\columnwidth}{!}{%
\begin{tabular}{|c|c c | c c l|} \hline 
Index
& \textbf{PS} 
& \textbf{MPO ($\Theta$)} 
& \textbf{Pin} 
& \textbf{\#Access} 
& \textbf{M1 Track Indices} \\ \hline 
\multirow{3}{*}{(a)}
&\multirow{3}{*}{--} 
& \multirow{3}{*}{--} 
& QN 
& $2$ 
& $1,2$ \\ 
& & & CLK 
& $2$ 
& $3, 4$ \\ 
& & 
& D 
& $3$ 
& $16, 19, 20$ \\ \hline 
\multirow{3}{*}{(b)}
& \multirow{3}{*}{$\checkmark$} 
& \multirow{3}{*}{--} 
& QN 
& $3$ 
& $1, 2, 3$ \\ 
& & & CLK 
& $4$ 
& $1,2,3,4$ \\ 
& & 
& D 
& $2$ 
& $19,20$ \\ \hline 
\multirow{3}{*}{(c)} 
& \multirow{3}{*}{$\checkmark$} 
& \multirow{3}{*}{4} 
& QN 
& $4$ 
& $1,2,3,4$ \\ 
& & 
& CLK 
& $7$ 
& $1,2,3,4,5,6,7$ \\ 
& & & D & $4$ & $15,16, 19,20$ \\ \hline 
\end{tabular}
}
\end{table}

\subsection{Effect of Acceleration Techniques}\label{subsec:accstudy}
\begin{figure}[ht]
\centering
\includegraphics[width=0.9\columnwidth]{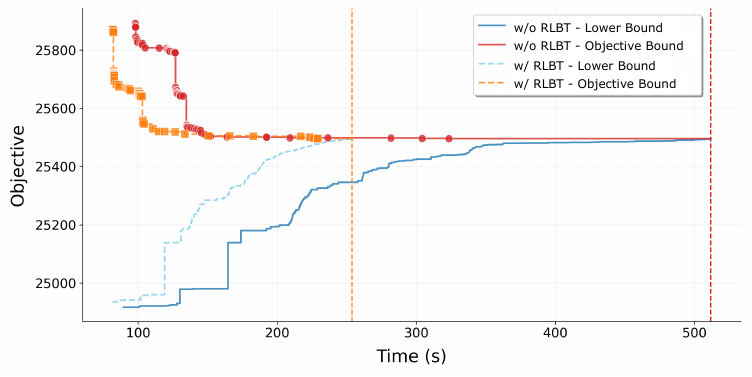}
\caption{CP-SAT search progress for DFFHQN\_X1 with and 
without routing lower bound tightening, showing objective 
and lower bounds over time. The vertical dashed 
line marks search termination.}
\label{fig:lbsearch}
\end{figure}

In this section, we evaluate the four 
acceleration techniques introduced in 
Section~\ref{sec:acc}. RLBT is always enabled in CPCell, 
as it directly reduces CP-SAT optimality 
certification overhead. As shown in \ywrone{Figure~\ref{fig:lbsearch}}, 
a tight routing lower bound substantially 
accelerates search convergence, reducing 
solver runtime from 512\,s to 254\,s (50.4\% speedup).

Table~\ref{tab:ablation_summary_acc} details the runtime impact of 
acceleration techniques on cells sourced from PROBE3.0~\cite{choi2023probe30}
(DFFHQN\_X1, DFFRNQ\_X1), ASAP7~\cite{asap7} (DFFHQN\_X3/X4, BUF\_X16), 
and SO3~\cite{so3cell2025} (2BDFFHQN\_X1) under 3:2 GR 
setting. A maximum runtime of 150000\,s is imposed, after which the best feasible 
solution is reported. \ywrone{Transistor clustering is deterministic given a fixed seed (in our studies, we use seed $= 42$). 
We default the parameters to $k_{\min}=2, k_{\max}=4$. If $k_{\max}=4$ reports infeasible, 
we manually lower the threshold by 1 until feasibility is met.} 
Transistor clustering consistently 
improves runtime while preserving optimal width, wirelength, and M2 usage. 
Specifically, DFFHQN\_X1 achieves a 36.6\% speedup with clustering ($k_{\max}=2$) 
and a 58.3\% speedup (106\,s) with clustering ($k_{\max}=4$). Similarly, DFFRNQ\_X1 
achieves a 42.1\% speedup with clustering ($k_{\max}=2$); however, aggressive clustering 
($k_{\max}=4$) renders the run unsatisfiable. For high-drive-strength cells, 
combining clustering with ITP yields the highest gains. DFFHQN\_X3 achieves a 
massive 88.9\% speedup (9211\,s to 1017\,s) using ITP and clustering ($k_{\max}=2$). 
DFFHQN\_X4 achieves a 47.5\% speedup using ITP and clustering ($k_{\max}=3$), though larger 
cluster sizes again lead to unsatisfiability. For large-scale designs, early 
termination via relative gap is necessary to avoid timeouts. 
A 0.5\% gap enables BUF\_X16 to converge in 533\,s with a reported cell width of 25 CPP.
A 1.0\% gap enables 2BDFFHQN\_X1 to converge in 15773\,s with a reported cell width of 25 CPP.
In summary, acceleration techniques significantly reduce search complexity 
but require careful tuning of cluster size and gap tolerance to maintain satisfiability.

\begin{table}[ht]
    \caption{Cell metrics and runtime for 3:2 GR cells. 
    Clst.: maximum cluster size $k_{\max}$. ITP: identical 
    transistor partitioning enablement. Gap: relative gap.}
    \label{tab:ablation_summary_acc}
    \setlength{\tabcolsep}{3.5pt} % Slightly increased for readability
    \centering
    \resizebox{\columnwidth}{!}{%
    \begin{tabular}{|l c c|c|c|c|c|c|c|r|}
        \hline
        \multirow{2}{*}{\textbf{Cell Name}}
        & \multirow{2}{*}{\textbf{\#FET}}
        & \multirow{2}{*}{\textbf{\#Net}}
        & \multicolumn{3}{c|}{\textbf{Acceleration}} 
        & \multirow{2}{*}{\textbf{\makecell{Width\\(CPP)}}}
        & \multirow{2}{*}{\textbf{\makecell{WL\\(nm)}}}
        & \multirow{2}{*}{\textbf{\#M2}}
        & \multirow{2}{*}{\textbf{\makecell{RT (s)}}} \\
        \cline{4-6}
        & & & \textbf{Clst.} & \textbf{ITP} & \textbf{Gap} & & & & \\
        \hline
        % DFFHQN_X1
        \multirow{3}{*}{\makecell{DFFHQN\_X1}} 
        & \multirow{3}{*}{24} & \multirow{3}{*}{17} 
        & --- & --- & --- & 14 & 2541.0 & 2 & 254 \\
        & & & 2 & --- & --- & 14 & 2541.0 & 2 & 161 \\
        & & & 4 & --- & --- & 14 & 2541.0 & 2 & \textbf{106} \\
        \hline
        % DFFRNQ_X1
        \multirow{3}{*}{\makecell{DFFRNQ\_X1}} 
        & \multirow{3}{*}{28} & \multirow{3}{*}{18} 
        & --- & --- & --- & 17 & 4166.0 & 4 & 11251 \\
        & & & 2 & --- & --- & 17 & 4166.0 & 4 & \textbf{6512} \\
        & & & 4 & --- & --- & --- & --- & --- & \emph{UNSAT} \\
        \hline
        % DFFHQN_X3
        \multirow{3}{*}{\makecell{DFFHQN\_X3}} 
        & \multirow{3}{*}{32} & \multirow{3}{*}{17} 
        & --- & --- & --- & 16 & 2826.5 & 2 & 9211 \\
        & & & 2 & \checkmark & --- & 16 & 2826.5 & 2 & \textbf{1017} \\
        & & & 3 & \checkmark & --- & --- & --- & --- & \emph{UNSAT} \\
        \hline
        % DFFHQN_X4
        \multirow{3}{*}{\makecell{DFFHQN\_X4}} 
        & \multirow{3}{*}{34} & \multirow{3}{*}{17} 
        & --- & --- & --- & 19 & 2966.5 & 2 & 133523 \\
        & & & 3 & \checkmark & --- & 19 & 2966.5 & 2 & \textbf{70156} \\
        & & & 4 & \checkmark & --- & --- & --- & --- & \emph{UNSAT} \\
        \hline
        % 2BDFFHQN_X1
        \multirow{3}{*}{\makecell{2BDFFHQN\_X1}} 
        & \multirow{3}{*}{44} & \multirow{3}{*}{29} 
        & --- & --- & --- & 25 & 5440.5 & 2 & \emph{Timeout} \\
        & & & 4 & --- & 0.5\% & 25 & 5060.5 & 2 & 146785 \\
        & & & 4 & --- & 1.0\% & 25 & 6042.0 & 2 & \textbf{15773} \\
        \hline
        % BUF_X16
        \multirow{3}{*}{\makecell{BUF\_X16}} 
        & \multirow{3}{*}{48} & \multirow{3}{*}{5} 
        & --- & --- & --- & 25 & 2876.0 & 0 & \emph{Timeout} \\
        & & & --- & \checkmark & --- & 25 & 2017.5 & 0 & \emph{Timeout} \\
        & & & --- & \checkmark & 0.5\% & 25 & 2876.0 & 0 & \textbf{533} \\
        \hline
    \end{tabular}
    }
\end{table}
% \vspace{1em}

\begin{table}[ht]
\caption{Cell metrics and runtime (RT) for different layout generation tools under 3:2 GR and 2F4T configuration for D flip-flops.}
\label{tab:ablation_summary_sota}
\centering
\setlength{\tabcolsep}{3pt}
\resizebox{\columnwidth}{!}{%
\begin{tabular}{|l c c|l|c|c|c|r|}
\hline
\textbf{Cell Name} 
& \textbf{\#FET} 
& \textbf{\#Net} 
& \textbf{\makecell{Tool}} 
& \textbf{\makecell{Width\\(CPP)}} 
& \textbf{\makecell{WL\\(nm)}}
& \textbf{\#M2} 
& \textbf{RT (s)} \\ \hline

\multirow{4}{*}{DFFHQN\_X1} & \multirow{4}{*}{24} & \multirow{4}{*}{17} 
  & PROBE3.0~\cite{choi2023probe30}   & 18 & 3803.0 & 2 & 17010 \\
& & & SMTCell-GR~\cite{agr} & 18 & 4504.5 & 2 & 9075 \\
& & & SO3~\cite{so3cell2025}        & 14 & 2577.0 & 2 & 262 \\
& & & Our work  & 14 & 2541.0 & 2 & 106 \\ \hline

\multirow{4}{*}{DFFRNQ\_X1} & \multirow{4}{*}{28} & \multirow{4}{*}{18} 
  & PROBE3.0~\cite{choi2023probe30}   & 20 & 5550.0 & 5 & 4576 \\
& & & SMTCell-GR~\cite{agr} & --- & --- & --- & \emph{Timeout} \\
& & & SO3~\cite{so3cell2025}       & 17 & 4197.0 & 4 & \emph{Timeout} \\
& & & Our work  & 17 & 4166.0 & 4 & 6512 \\ \hline

\multirow{4}{*}{2BDFFHQN\_X1} & \multirow{4}{*}{44} & \multirow{4}{*}{29} 
  & PROBE3.0~\cite{choi2023probe30}   & --- & --- & --- & \emph{Timeout} \\
& & & SMTCell-GR~\cite{agr} & --- & --- & --- & \emph{Timeout} \\
& & & SO3~\cite{so3cell2025}        & --- & --- & --- & \emph{Timeout} \\
& & & Our work  & 25 & 6042.0 & 2 & 15773 \\ \hline

\end{tabular}
}
\end{table}

\subsection{\ywrone{Scalability} Comparison with Other Tools for PROBE3.0}\label{subsec:sotastudy}
Table~\ref{tab:ablation_summary_sota} compares layout generation tools for D flip-flop designs using 3:2 GR against the best 
runtime results from Table~\ref{tab:ablation_summary_acc}. All tools were 
tested on the same machine with a 150000\,s 
timeout, using their respective acceleration techniques. Specifically, 
PROBE3.0 and SMTCell-GR use partitioned netlists from prior work~\cite{spr} 
and the Z3 solver~\cite{SMT} with multi-threading. SO3~\cite{so3cell2025} 
uses Gurobi~\cite{gurobi} with 12 workers (double our CP-SAT configuration) 
and restricts topology optimization, gate alignments, diffusion breaks, and 
occupying track to reduce runtime. 
Given variations in rule support and encoding across tools, width serves 
as the primary metric, with wirelength and M2 track count provided for reference. 
CPCell requires no prior solution 
data, allowing seamless application to unseen netlists. It consistently achieves 
the smallest (or tied) cell width (CPP) often with lower runtime.

\section{Block-level Evaluations on Gear-ratio-enabled Cell Layouts}
\label{sec:block-exp}

In this section, we explore example impacts of gear ratio 
at block-level. 
% We use 45:45 (1:1) GR, 45:45+22.5 (1:1+$\frac{1}{2}$) GR, 45:30 (3:2) GR, 45:27 (5:3) GR
First, we confirm whether using offset cell layouts can 
in practice mitigate the placement legalization bottleneck. 
Second, we perform a comparative analysis of GR choices in terms 
of block-level PPA. 
Third, we evaluate IR-drop to show power integrity trade-offs 
introduced by GR alternatives.
% Third, we evaluate IR-drop to capture power-integrity trade-offs introduced by tighter GR.

These examples show how our framework can illuminate such phenomena as
the following:
\begin{enumerate}[label=(\arabic*)]
    \item Offset cell layouts can improve the block PPA while also reducing
    runtime.
    \item Although a smaller M1 pitch generally benefits 
    block-level routing, it does not always lead to better PPA because of 
    the trade-off between routing resources and wire capacitances.
    \item A denser power distribution network (PDN) enabled by a GR with 
    a smaller M1 pitch does not necessarily lead to smaller IR-drop 
    compared to the sparser PDN of a GR with a larger M1 pitch.
    \setcounter{hypo}{\value{enumi}}
\end{enumerate}

Standard cell LVS and parasitic extraction are implemented with 
Cadence Pegasus v21.3~\cite{pegasus} and Quantus v21.1~\cite{quantus}, and characterization is
performed with Cadence Liberate v23.1~\cite{liberate}. \ywrone{Cell layout signoff is implemented with KLayout~\cite{klayout}.}
Synthesis and place-and-route (P\&R) are implemented with 
Synopsys Design Compiler \ywrone{R-2020.09}~\cite{designcompiler} and IC Compiler II \ywrone{R-2020.09}~\cite{icc2}, respectively. 
IR-drop is measured with Cadence Voltus v21.1~\cite{voltus}.
P\&R and IR-drop are executed using 16 threads on a server equipped
with an Intel Xeon Gold 6148 CPU (2.40 GHz, 80 threads) and 384 GB of memory. Our scripts are available at \cite{cpcell}.
% \ywrone{All layout signed off with KLayout~\cite{klayout}

%Finally, we present a case study on using the ECG flow to further improve a block design. 

\subsection{Gear Ratio \& Offset Enablement}
Prior work~\cite{agr} identifies block-level legalization failure and
proposes offset cells as a remedy. We evaluate 1:1, 3:2, and 5:3 GR
libraries. For 3:2 GR, 15\,nm-offset cells (3:$2^{\dagger}$) are grouped
with 0-offset cells in an equivalent cell group (ECG).\footnote{Equivalent
Cell Groups contain functionally identical physical variants, e.g., with
different pin locations or colors. Variant-aware legalization swaps a
misaligned cell for an aligned variant instead of moving it.}
For 5:3 GR, 9\,nm- and 18\,nm-offset cells (5:$3^{\dagger}$ and
5:$3^{\dagger\dagger}$) are similarly grouped with 0-offset cells.

\begin{figure}[ht]
	\centering
	\includegraphics[width=1.0\columnwidth]{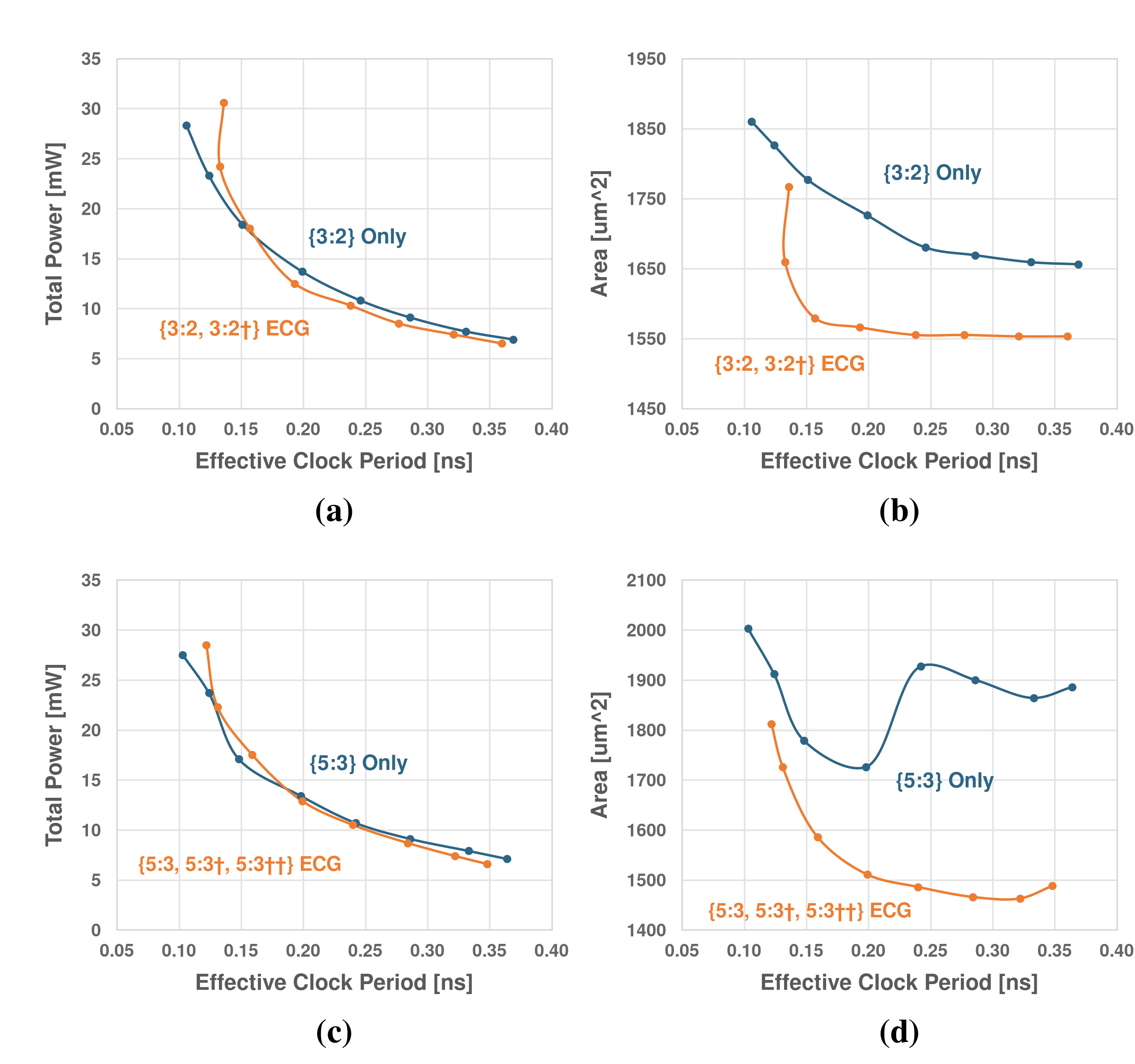}
	\caption{Offset-cell PPA: (a) 3:2 GR total power; (b) 3:2 GR area;
(c) 5:3 GR total power; (d) 5:3 GR area, each vs. effective CLKP.}
	\label{fig:eval_offset}
\end{figure}

Table~\ref{tab:block_eval_data} shows data for
\ywrone{JPEG}~\cite{jpeg} across target clock periods (TCPs) from 
100\,ps to 400\,ps. Performance is represented by the effective clock period (CLKP):
TCP minus worst negative slack.
\ywrone{For each nominal TCP, we run the flow at TCP, TCP 
$\pm\,0.1$ ps and TCP $\pm\,0.2$ ps \textcolor{black}{over a range of utilizations. 
At each utilization, runs that fail to route or have \#DRV$\;>300$ are discarded. 
We select the highest evaluated utilization at which at least one of the 
five runs remains, and report the median of each metric over all 
remaining runs. 
The medians are computed per metric and do not necessarily correspond to a 
single run, so the comparisons in Table~\ref{tab:block_eval_data} show 
marginal trends across GR options. 
All evaluated runs are listed in our 
repository~\cite{cpcell}.}
Each TCP is evaluated independently, and failure at one TCP is
not assumed to imply failure at any other TCP. We compute the
effective clock period as $\mathrm{TCP}-\mathrm{WNS}$, where WNS is the worst negative slack.}
Figure~\ref{fig:eval_offset} shows the PPA comparison plot for 3:2 and 5:3 GR.
For both gear ratios, adding offset variants reduces area with similar 
power and performance, except at the 100\,ps TCP.
%This supports the first hypothesis.

\begin{figure}[ht]
	\centering
    \includegraphics[width=1.0\columnwidth]{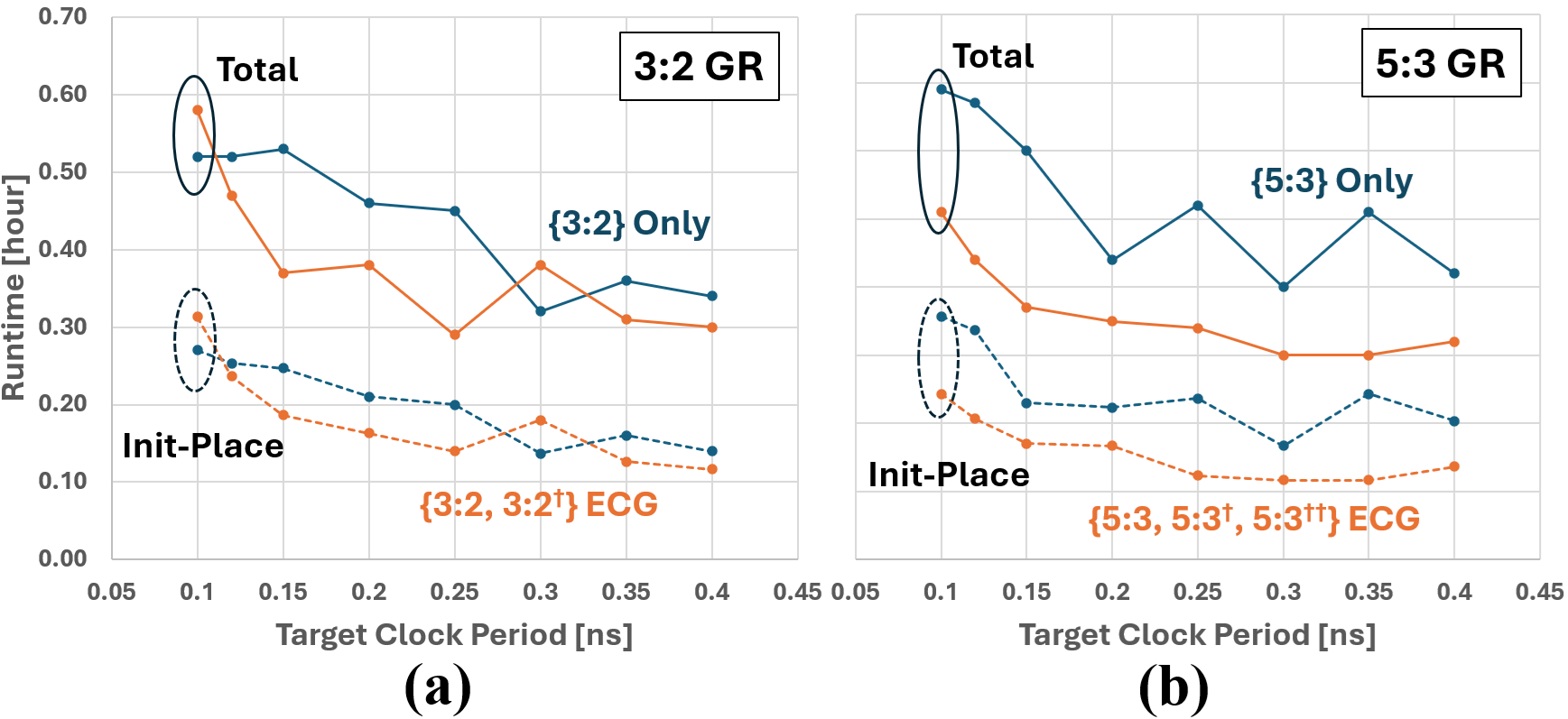}
	\caption{Block-level P\&R runtime comparison. 
 (a) 3:2 GR; (b) 5:3 GR.}
	\label{fig:runtime}
 % \vspace{0.1cm}
\end{figure}

Figure~\ref{fig:runtime} shows P\&R runtime comparison results. 
The runtime data confirm that 
ECGs yield shorter runtimes than 0-offset sets in most cases, 
possibly suggesting that finding legalizable physical variants takes less time 
than finding legalizable placement locations.

% \vspace{-0.15in}

\begin{table*}[th]
\centering
\scriptsize
\caption{Comparison of Block-Level Metrics across Target Clock Periods}
\renewcommand{\arraystretch}{1.0}      % vertical height
\setlength{\tabcolsep}{12.5pt}          % horizontal height
\label{tab:block_eval_data}
\resizebox{\textwidth}{!}{%
\begin{tabular}{|c|c|l|l|r|r|r|r|r|r|r|r|}
\hline
\multicolumn{4}{|c|}{\multirow{2}{*}{}} & \multicolumn{8}{c|}{\textbf{Target Clock Period (ns)}} \\ \cline{5-12} 
\multicolumn{4}{|c|}{} & 0.10 & 0.12 & 0.15 & 0.20 & 0.25 & 0.30 & 0.35 & 0.40 \\ \hline

% ================= 1:1 GR BLOCK =================
\multicolumn{2}{|c|}{\multirow{6}{*}{1:1 GR}} & \multicolumn{2}{l|}{\#Insts} & 60255 & 47718 & 44714 & 43774 & 43215 & 42965 & 42850 & 42700 \\ \cline{3-12} 
\multicolumn{2}{|c|}{} & \multicolumn{2}{l|}{Wirelength ($\mu \mathrm{m}$)} & 92560 & 87659 & 82759 & 73234 & 69993 & 70875 & 68578 & 69116 \\ \cline{3-12} 
\multicolumn{2}{|c|}{} & \multicolumn{2}{l|}{Total Power (mW)} & 29.7 & 24.6 & 18.2 & 13.0 & 10.3 & 8.9 & 7.6 & 6.7 \\ \cline{3-12} 
\multicolumn{2}{|c|}{} & \multicolumn{2}{l|}{Worst Negative Slack (ns)} & -0.021 & -0.013 & 0.000 & 0.009 & 0.014 & 0.022 & 0.028 & 0.034 \\ \cline{3-12} 
\multicolumn{2}{|c|}{} & \multicolumn{2}{l|}{Effective CLKP (ns)} & 0.121 & 0.133 & 0.150 & 0.191 & 0.236 & 0.278 & 0.322 & 0.366 \\ \cline{3-12} 
\multicolumn{2}{|c|}{} & \multicolumn{2}{l|}{Area ($\mu \mathrm{m}^2$)} & 2199 & 2017 & 1935 & 1886 & 1878 & 1836 & 1838 & 1836 \\ \hline\hline

% ================= 3:2 GR BLOCK =================
\multirow{12}{*}{3:2 GR} & \multirow{6}{*}{\shortstack{0-offset\\Only}} & \multicolumn{2}{l|}{\#Insts} & 57889 & 54403 & 52951 & 47918 & 46307 & 45750 & 45319 & 45376 \\ \cline{3-12} 
 & & \multicolumn{2}{l|}{Wirelength ($\mu \mathrm{m}$)} & 74528 & 71986 & 71602 & 72681 & 65940 & 64729 & 63707 & 64154 \\ \cline{3-12} 
 & & \multicolumn{2}{l|}{Total Power (mW)} & 28.3 & 23.3 & 18.4 & 13.7 & 10.8 & 9.1 & 7.7 & 6.9 \\ \cline{3-12} 
 & & \multicolumn{2}{l|}{Worst Negative Slack (ns)} & -0.006 & -0.004 & -0.001 & 0.001 & 0.004 & 0.014 & 0.019 & 0.031 \\ \cline{3-12} 
 & & \multicolumn{2}{l|}{Effective CLKP (ns)} & 0.106 & 0.124 & 0.151 & 0.199 & 0.246 & 0.286 & 0.331 & 0.369 \\ \cline{3-12} 
 & & \multicolumn{2}{l|}{Area ($\mu \mathrm{m}^2$)} & 1860 & 1826 & 1777 & 1726 & 1680 & 1669 & 1659 & 1656 \\ \cline{2-12} 

 & \multirow{6}{*}{\shortstack{Mixed-\\offset}} & \multicolumn{2}{l|}{\#Insts} & 57155 & 44195 & 42117 & 41543 & 41207 & 41103 & 41106 & 41087 \\ \cline{3-12} 
 & & \multicolumn{2}{l|}{Wirelength ($\mu \mathrm{m}$)} & 76773 & 79973 & 66652 & 59613 & 56396 & 55492 & 57627 & 55264 \\ \cline{3-12} 
 & & \multicolumn{2}{l|}{Total Power (mW)} & 30.6 & 24.2 & 18.0 & 12.5 & 10.3 & 8.5 & 7.4 & 6.5 \\ \cline{3-12} 
 & & \multicolumn{2}{l|}{Worst Negative Slack (ns)} & -0.036 & -0.013 & -0.007 & 0.007 & 0.012 & 0.023 & 0.029 & 0.040 \\ \cline{3-12} 
 & & \multicolumn{2}{l|}{Effective CLKP (ns)} & 0.136 & 0.133 & 0.157 & 0.193 & 0.238 & 0.277 & 0.321 & 0.360 \\ \cline{3-12} 
 & & \multicolumn{2}{l|}{Area ($\mu \mathrm{m}^2$)} & 1767 & 1659 & 1579 & 1566 & 1555 & 1555 & 1553 & 1553 \\ \hline\hline

% ================= 5:3 GR BLOCK =================
\multirow{12}{*}{5:3 GR} & \multirow{6}{*}{\shortstack{0-offset\\Only}} & \multicolumn{2}{l|}{\#Insts} & 59183 & 57455 & 52756 & 47062 & 54266 & 53125 & 52780 & 52663 \\ \cline{3-12} 
 & & \multicolumn{2}{l|}{Wirelength ($\mu \mathrm{m}$)} & 77686 & 80350 & 68864 & 73755 & 72539 & 70100 & 70454 & 68210 \\ \cline{3-12} 
 & & \multicolumn{2}{l|}{Total Power (mW)} & 27.5 & 23.7 & 17.1 & 13.4 & 10.7 & 9.1 & 7.9 & 7.1 \\ \cline{3-12} 
 & & \multicolumn{2}{l|}{Worst Negative Slack (ns)} & -0.003 & -0.004 & 0.002 & 0.003 & 0.008 & 0.014 & 0.017 & 0.036 \\ \cline{3-12} 
 & & \multicolumn{2}{l|}{Effective CLKP (ns)} & 0.103 & 0.124 & 0.148 & 0.198 & 0.242 & 0.286 & 0.333 & 0.364 \\ \cline{3-12} 
 & & \multicolumn{2}{l|}{Area ($\mu \mathrm{m}^2$)} & 2003 & 1912 & 1779 & 1726 & 1927 & 1900 & 1864 & 1886 \\ \cline{2-12}

 & \multirow{6}{*}{\shortstack{Mixed-\\offset}} & \multicolumn{2}{l|}{\#Insts} & 54104 & 50761 & 41131 & 39900 & 39594 & 38890 & 38803 & 38731 \\ \cline{3-12} 
 & & \multicolumn{2}{l|}{Wirelength ($\mu \mathrm{m}$)} & 72021 & 68720 & 71780 & 66803 & 64239 & 57964 & 55328 & 55335 \\ \cline{3-12} 
 & & \multicolumn{2}{l|}{Total Power (mW)} & 28.5 & 22.3 & 17.5 & 12.9 & 10.5 & 8.7 & 7.4 & 6.6 \\ \cline{3-12} 
 & & \multicolumn{2}{l|}{Worst Negative Slack (ns)} & -0.022 & -0.011 & -0.009 & 0.001 & 0.010 & 0.016 & 0.028 & 0.052 \\ \cline{3-12} 
 & & \multicolumn{2}{l|}{Effective CLKP (ns)} & 0.122 & 0.131 & 0.159 & 0.199 & 0.240 & 0.284 & 0.322 & 0.348 \\ \cline{3-12} 
 & & \multicolumn{2}{l|}{Area ($\mu \mathrm{m}^2$)} & 1812 & 1726 & 1586 & 1511 & 1486 & 1466 & 1463 & 1488 \\ \hline
\end{tabular}%
} 
\end{table*}

\subsection{Choice of Optimal GR in DTCO Context}

\begin{figure}[ht]
	\centering
	\includegraphics[width=1.0\columnwidth]{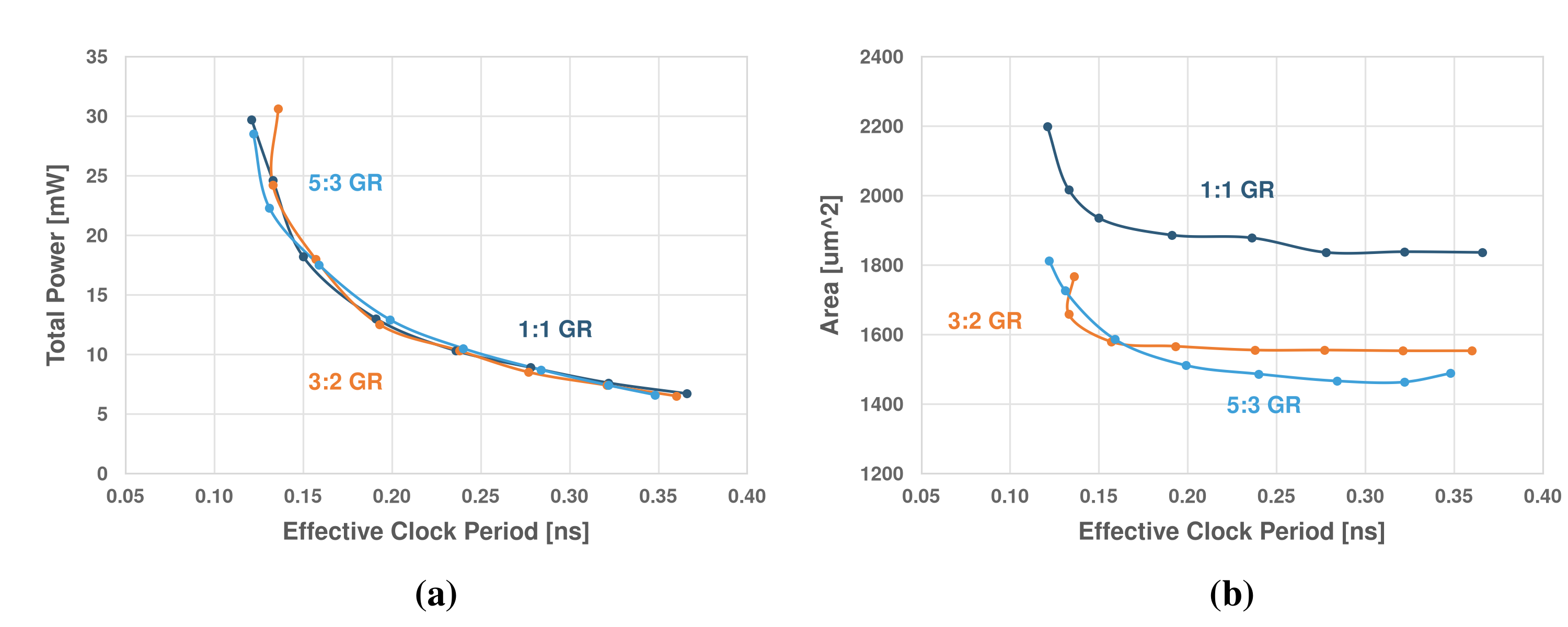}
	\caption{PPA comparison for identifying the optimal GR in the DTCO context.
 (a) total power vs. effective clock period; (b) area vs. effective clock period.}
	\label{fig:eval_GR}
\end{figure}

Smaller M1 pitch reduces routing congestion but increases wire
capacitance and power, making GR selection a PPA trade-off for DTCO.
Figure~\ref{fig:eval_GR} compares GRs\footnote{ECGs are used for 3:2 and
5:3 GRs.} and shows area benefits for 3:2 GR with power and performance
comparable to 1:1 and 5:3 GRs, except at the 100\,ps TCP.

\begin{enumerate}[label=(\arabic*)]
    % \item The three GR show equivalent power and performance in the low-frequency
    % range. However, they diverge at high-frequency range, where the 1:1 GR exhibits
    % slightly better power and performance. 
    \item Figure~\ref{fig:eval_GR}(a) indicates that power consumption is 
    comparable across all GRs over the entire frequency range.
    % \item In the area-performance plot, the 3:2 and 5:3 GR show a clear area benefit
    % over the 1:1 GR.
    \item Figure~\ref{fig:eval_GR}(b) indicates that the 3:2 and 5:3 GRs 
    provide significant area reductions compared to the 1:1 GR across the 
    entire frequency range.
    % \item The 5:3 GR shows an area benefit over the 3:2 GR in the low-frequency 
    % range, while the 3:2 GR shows area benefit over the 5:3 GR in the high-frequency 
    % range.
    \item Figure~\ref{fig:eval_GR}(b) reveals a crossover in area efficiency: 
    the 5:3 GR requires the least area at lower frequencies, whereas the 
    3:2 GR becomes more area-efficient in the high-frequency regime.
\end{enumerate}

These studies suggest that selection of the best GR option
may depend on the target design parameters.

\subsection{Impact of Gear Ratio on IR-drop}
\begin{figure}[ht]
	\centering
	\includegraphics[width=0.85\columnwidth]{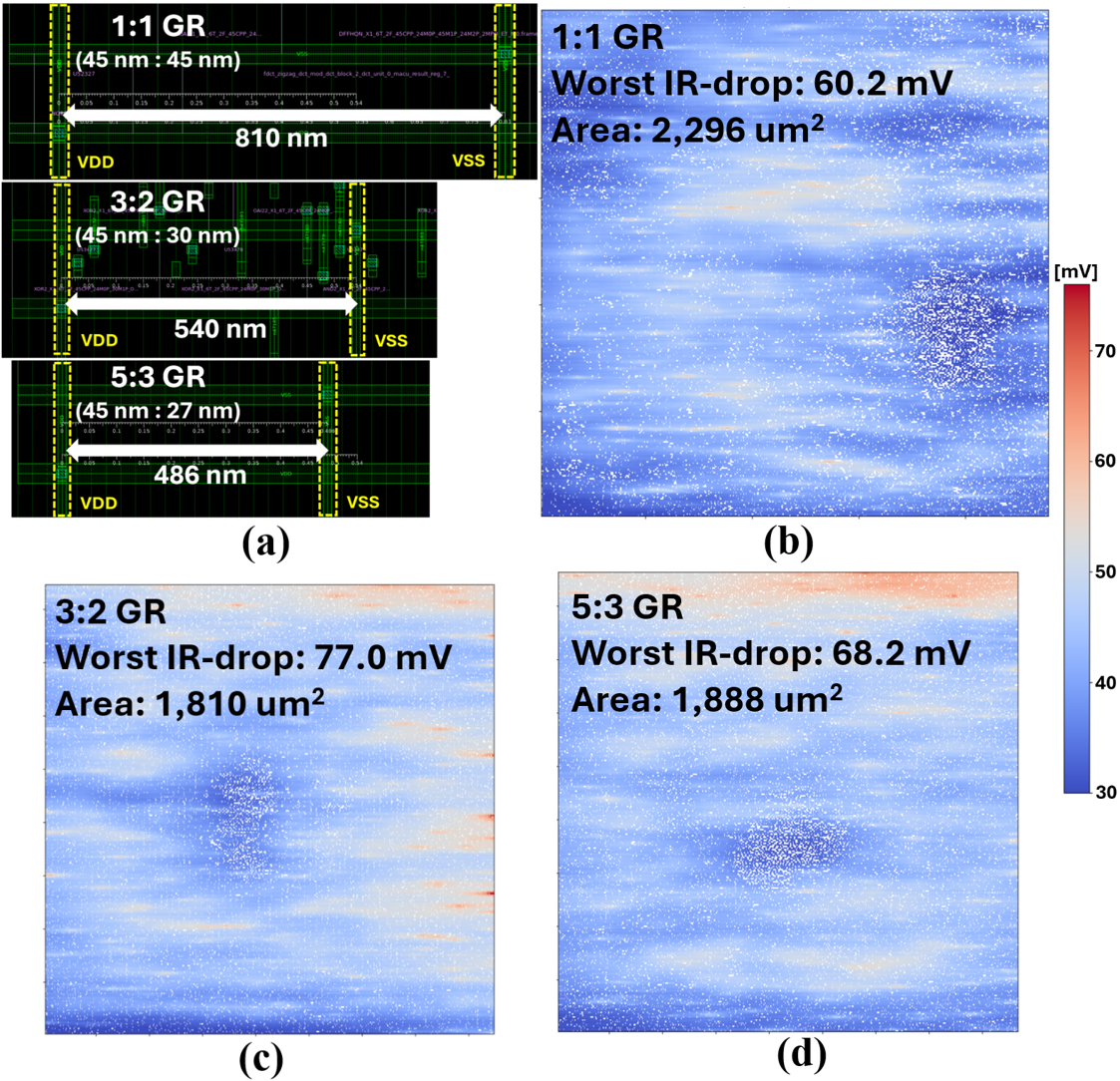}
	\caption{GR-dependent static IR-drop with M1-pitch-scaled power stripe
spacing: (a) stripe spacing; IR-drop maps for (b) 1:1, (c) 3:2, and
(d) 5:3 GR.}
	\label{fig:eval_IR}
 % \vspace{0.1cm}
\end{figure}

Gear ratio is used not only for PPA benefit but also for IR-drop 
reduction.
With a smaller M1 pitch, additional routing resources on M1 can be 
used to enable a finer PDN mesh to mitigate IR-drop. 
Figure~\ref{fig:eval_IR}(a) shows the setting of power stripe distance 
for each GR to 18 times the M1 pitch, which effectively places 
the worst IR-drop at approximately 10\% of VDD at a clock period of 100\,ps: 
810\,nm, 540\,nm and 486\,nm for the 1:1, 3:2 and 5:3 GR, respectively. 
Figure~\ref{fig:eval_IR} also shows that the 3:2 GR has the worst IR-drop 
(77.0\,mV), while the 1:1 GR has the lowest IR-drop (60.2\,mV). 
The 3:2 and 5:3 GRs show a higher IR-drop than 
the 1:1 GR despite their denser PDNs, 
because their smaller areas induce higher current densities.
From such IR-drop evaluation studies, a technology developer 
could, e.g., decide to apply the 5:3 GR to gain IR-drop 
benefits, accepting the higher cost required for denser M1 patterning.
This highlights the necessity of holistic block-level 
evaluation in DTCO decision-making.

\subsection{\ywrone{Ablation Study on DTCO Features}}\label{subsec:dtcoablation}
\ywrone{We quantify the block-level contribution of two DTCO features: the M0 pin accessibility constraints of
Section~\ref{sec:m0pin} and parameterizable gate cuts. Both studies use
\ywrone{JPEG}~\cite{jpeg} at 0.90 utilization and a target clock
period of 0.1\,ns.}

\ywrone{We first assess downstream routability across configurations (a),
(b) and (c) of Figure~\ref{fig:pinaccess}. Configuration (a) applies no pin
access constraint, (b) applies pin separation (PS), and (c) applies both PS
and minimum pin opening (MPO). Configurations (a) and (b) yield identical
average cell width, while (c) widens the congested cell XOR2\_X1 by one CPP
to secure additional pin opening. Table~\ref{tab:pin_ppa} reports
block-level wirelength, power, timing, area and post-route design rule
violation count (\#DRV). PS reduces the reported post-route DRV count from 
277 to 11 without changing block area. With PS and MPO, the reported count 
is zero. Relative to the unconstrained library, the combined configuration 
increases block area by 1.7\% and reported total power by 4.9\%. However, improved routability is observed in the evaluated JPEG configuration.}

\ywrone{We quantify the contribution of enabling gate cuts.
Table~\ref{tab:cut_ppa} compares configuration (c) with and without the
2-CPP gate cut. The gate cut reduces average cell width by 3.0\%
(6.65 to 6.45 CPP), with the reduction concentrated in cells whose transistor
netlists contain crossings such as MUX2\_X1, LHQ\_X1 and DFFHQN\_X1. At the
block level this yields 6.5\% area reduction
(2482 to 2320\,$\mu$m$^2$) and 3.3\% wirelength reduction, at a cost of
1.3\% power increase and an added 1\,ps of WNS.}

\begin{table}[tb]
\centering
\scriptsize
\ywrone{
\caption{Pin-accessibility metrics for \textnormal{(a)} PA\_A,
\textnormal{(b)} PA\_B, and \textnormal{(c)} PA\_C
(Figure~\ref{fig:pinaccess}), at $0.90$ utilization and
$\mathrm{TCP}=0.10\,\mathrm{ns}$. Values are rounded to two decimal places.}
\label{tab:pin_ppa}
\setlength{\tabcolsep}{3pt}
\resizebox{0.8\columnwidth}{!}{
\begin{tabular}{|l|r|r|r|}
\hline
\multirow{2}{*}{Metric} &
\multicolumn{3}{c|}{Pin-Accessibility Configuration} \\ \cline{2-4}
& \makebox[1.05cm][c]{(a)} & \makebox[1.05cm][c]{(b)} & \makebox[1.05cm][c]{(c)} \\ \hline
Avg. Cell Width (CPP)
& 6.43 & 6.43 & 6.45 \\ \hline
Wirelength ($\mu\mathrm{m}$)
& 94964 & 94503 & 96273 \\ \hline
Total Power (mW)
& 28.7 & 30.0 & 30.1 \\ \hline
Worst Negative Slack (ns)
& -0.005 & -0.003 & -0.010 \\ \hline
Effective CLKP (ns)
& 0.105 & 0.103 & 0.110 \\ \hline
Area ($\mu\mathrm{m}^{2}$)
& 2282 & 2282 & 2320 \\ \hline
\#DRV
& 277 & 11 & 0 \\ \hline
\end{tabular}
}
}
\end{table}

\begin{table}[tb]
\centering
\scriptsize
\ywrone{
\caption{Cell-level and block-level metrics with or without gate cut. Block-level results use 0.90 utilization and $\mathrm{TCP} = 0.10\,\mathrm{ns}$.}
\label{tab:cut_ppa}
\setlength{\tabcolsep}{6pt}
\resizebox{0.9\columnwidth}{!}{
\begin{tabular}{|l|*{2}{>{\raggedleft\arraybackslash}p{2.0cm}|}}
\hline
\multirow{2}{*}{Metric} & \multicolumn{2}{c|}{Gate Cut Setting}
 \\ \cline{2-3}
 & \multicolumn{1}{c|}{w/o Gate Cut } & \multicolumn{1}{c|}{w/ Gate Cut (2-CPP)} \\ \hline
Avg. Cell Width (CPP) & 6.65 & 6.45 \\ \hline
Wirelength ($\mu \mathrm{m}$) & 99538 & 96273 \\ \hline
Total Power (mW) & 29.7 & 30.1 \\ \hline
Worst Negative Slack (ns) & -0.009 & -0.010 \\ \hline
Effective CLKP (ns) & 0.109 & 0.110\\ \hline
Area ($\mu \mathrm{m}^2$) & 2482 & 2320 \\ \hline
\end{tabular}
}
}
\end{table}

\section{Conclusion}\label{sec:conclusion}
CPCell extends gear-ratio studies with a standard-cell layout framework
supporting arbitrary gear ratios and offset variants for DTCO under
PROBE3.0. Its acceleration techniques reduce CP-SAT runtime by up to
88.96\%, enabling generation for netlists with up to 48 transistors.
Cell-level experiments show reduced average width for 3:2 and 5:3 GRs,
with cell-dependent wirelength changes. Block-level results show area
and wirelength benefits, with PPA and IR-drop trade-offs that depend on
target conditions. CPCell connects cell optimization to block implementation
for systematic technology and architecture evaluation.

\section*{Acknowledgment}
The authors acknowledge the use of ChatGPT 
for improving the English phrasing and grammatical 
 quality of the manuscript.

\bibliographystyle{IEEEtran}
\begin{footnotesize}
\addtolength{\baselineskip}{-1pt}

\end{footnotesize}
\begin{IEEEbiography}[{\raisebox{0.3in}{\includegraphics[height=0.8in,clip,keepaspectratio]
{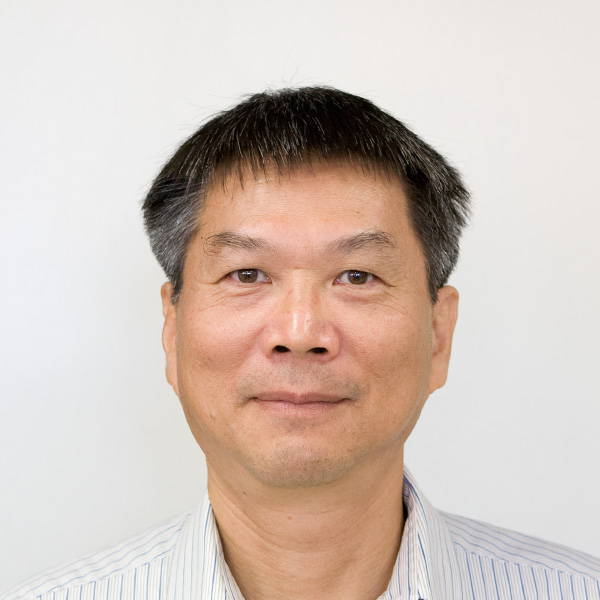}}}]{Chung-Kuan Cheng}
is Distinguished Professor of CSE and Adjunct Professor of
ECE at the University of California, San Diego.
His research interests include machine learning 
and design automation for microelectronic circuits. He received the Ph.D. degree 
in Electrical Engineering and Computer Sciences from the University of
California, Berkeley.
\end{IEEEbiography}

\vspace{-0.9in}

\begin{IEEEbiography}
[{\raisebox{0.4in}{\includegraphics[height=0.8in, clip, keepaspectratio]
{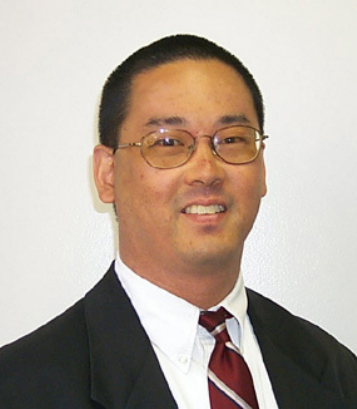}}}]
{Andrew B. Kahng} is Distinguished Professor of CSE and ECE at the 
University of California, San Diego. His interests include IC physical design, 
the design-manufacturing interface, combinatorial optimization, 
and AI/ML for EDA and IC design. He received the Ph.D. degree in Computer Science from the University of California, San Diego.
\end{IEEEbiography}

\vspace{-0.9in}

\begin{IEEEbiography}[{\raisebox{0.4in}{\includegraphics[height=0.8in,clip,keepaspectratio]{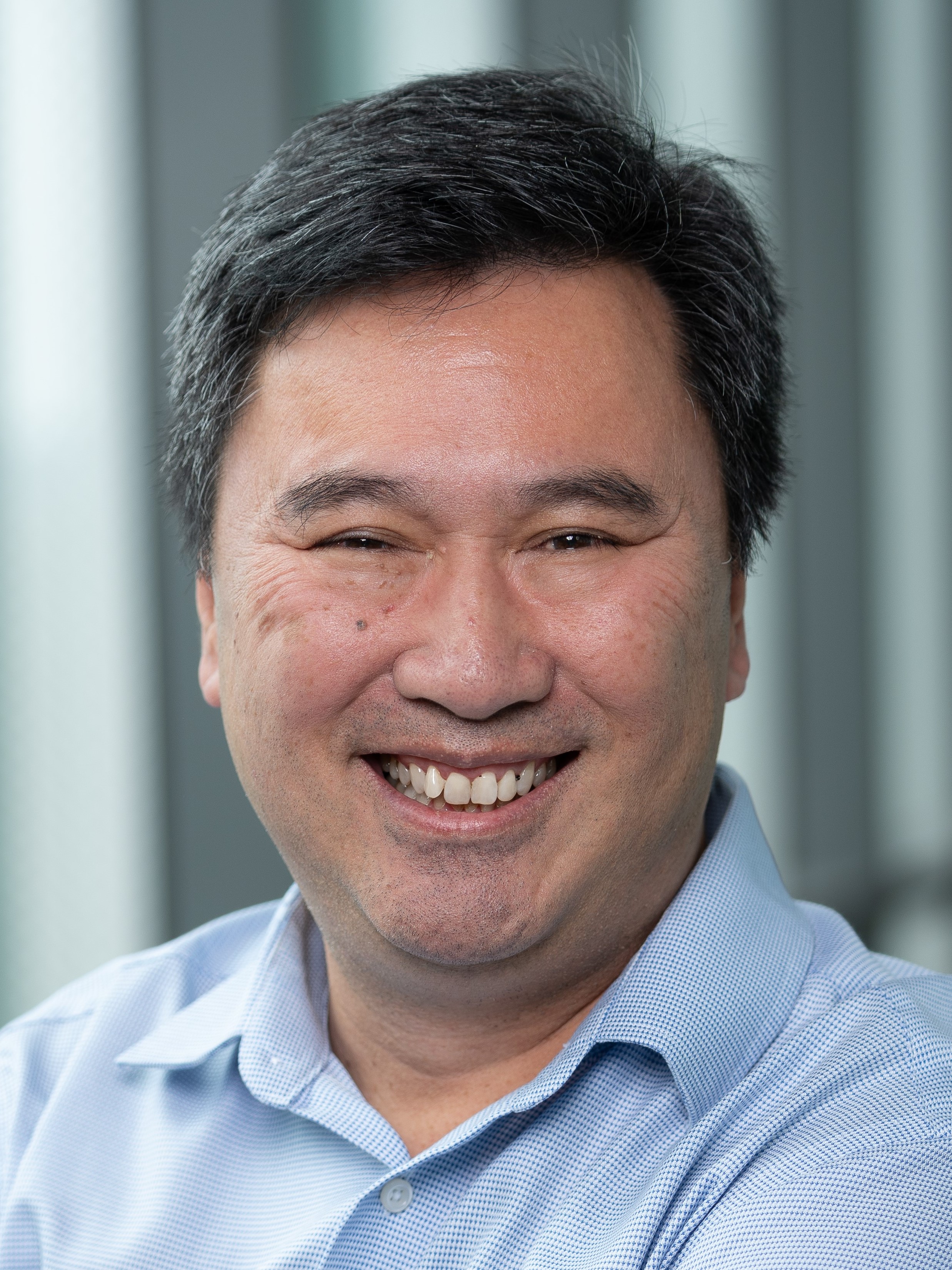}}}]{Bill Lin} received BS, MS and PhD degrees in EECS from UC Berkeley. He is a Professor of Electrical and Computer Engineering at UC San Diego, affiliated with the Center for Wireless Communications and the Center for Networked Systems, CNS. He is also currently the Associate Dean for Research for the Jacobs School of Engineering. 
\end{IEEEbiography}
% He has over 250 publications, multiple best-paper awards, six patents and extensive IEEE/ACM service.  

\vspace{-0.9in}

\begin{IEEEbiography}
[{\raisebox{0.4in}{\includegraphics[height=0.8in, clip, keepaspectratio]{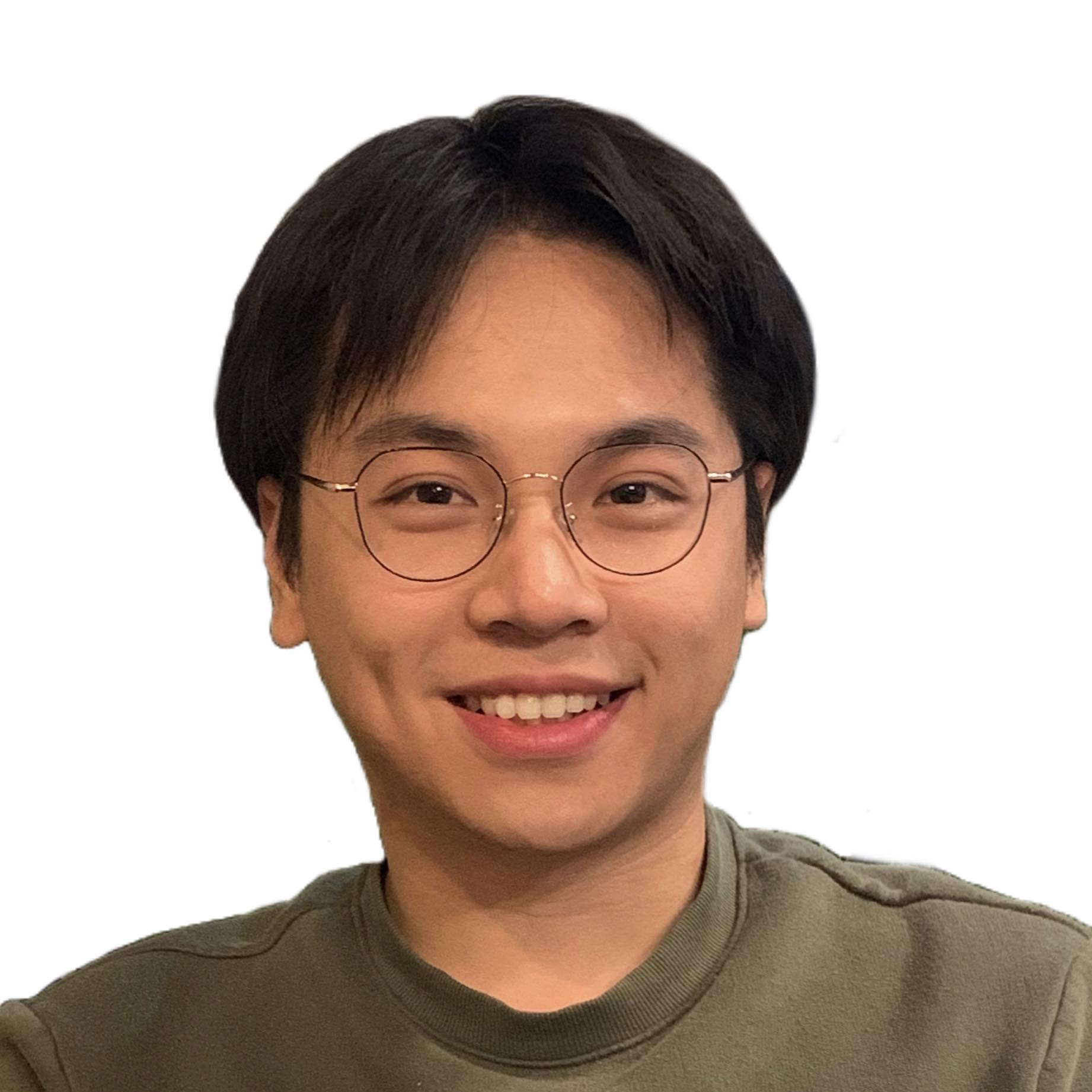}}}]
{Yucheng Wang} is a Ph.D. student in the Computer Science Department of the University of California at San Diego. He received his bachelor's degree in Computer Science from Purdue University, West Lafayette, in 2021. His research interests include standard-cell layout automation, design technology co-optimization strategies and graph visualization.
\end{IEEEbiography}

\vspace{-0.9in}

\begin{IEEEbiography}
[{\raisebox{0.4in}{\includegraphics[height=0.8in, clip, keepaspectratio]{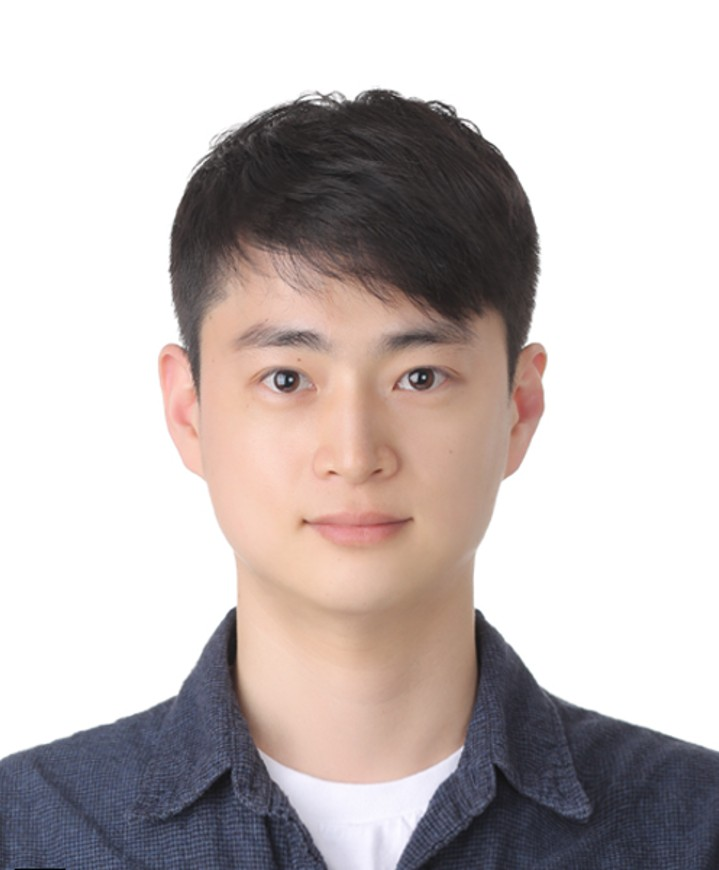}}}]
{Dooseok Yoon} received the M.S. degree in electrical and computer engineering
from the University of California at San Diego, La Jolla, CA,
USA, in 2024. He is currently pursuing the Ph.D. degree 
at the University of California at San Diego, La Jolla.
His research interests include VLSI physical design and 
design-technology co-optimization.
\end{IEEEbiography}

% \newpage

% \section{Biography Section}
% If you have an EPS/PDF photo (graphicx package needed), extra braces are
%  needed around the contents of the optional argument to biography to prevent
%  the LaTeX parser from getting confused when it sees the complicated
%  $\backslash${\tt{includegraphics}} command within an optional argument. (You can create
%  your own custom macro containing the $\backslash${\tt{includegraphics}} command to make things
%  simpler here.)
 
% \vspace{11pt}

% % \bf{If you include a photo:}\vspace{-33pt}
% % \begin{IEEEbiography}[{\includegraphics[width=1in,height=1.25in,clip,keepaspectratio]{fig1}}]{Michael Shell}
% % Use $\backslash${\tt{begin\{IEEEbiography\}}} and then for the 1st argument use $\backslash${\tt{includegraphics}} to declare and link the author photo.
% % Use the author name as the 3rd argument followed by the biography text.
% % \end{IEEEbiography}

% \vspace{11pt}

% \bf{If you will not include a photo:}\vspace{-33pt}
% \begin{IEEEbiographynophoto}{John Doe}
% Use $\backslash${\tt{begin\{IEEEbiographynophoto\}}} and the author name as the argument followed by the biography text.
% \end{IEEEbiographynophoto}

% \vfill

\end{document}